\documentclass[reprint,aip,jap]{revtex4-1}
\usepackage{graphicx, dcolumn, bm, xcolor}
\usepackage{textcomp,float}
\usepackage{amsmath}
\usepackage{natbib}
\usepackage{ulem} 
\newcommand{\beginsupplement}{%
        \setcounter{table}{0}
        \renewcommand{\thetable}{S\arabic{table}}%
        \setcounter{figure}{0}
        \renewcommand{\thefigure}{S\arabic{figure}}%
     }

\begin{document}
\title{On a critical artifact in the quantum yield methodology}

\author{Bart van Dam}
\affiliation{University of Amsterdam, Institute of Physics, Science Park 904, 1098 XH Amsterdam, The Netherlands}
\author{Benjamin Bruhn}
\affiliation{University of Amsterdam, Institute of Physics, Science Park 904, 1098 XH Amsterdam, The Netherlands}
\author{Ivo Kondapaneni}
\affiliation{Computer Graphics Group, KSVI, Faculty of Mathematics and Physics, Charles University, Malostransk\'{e} n\'{a}m\v{e}st\'{i} 25, Prague 1 CZ-11800, Czechia}
\author{Gejza Dohnal}
\affiliation{Czech Technical University in Prague, Faculty of Mechanical Engineering, Karlovo n\'{a}m\v{e}st\'{i} 13, 121 35, Prague 2, Czech Republic  }
\author{Alexander Wilkie}
\affiliation{Computer Graphics Group, KSVI, Faculty of Mathematics and Physics, Charles University, Malostransk\'{e} n\'{a}m\v{e}st\'{i} 25, Prague 1 CZ-11800, Czechia}
\author{Jaroslav K\v{r}iva\'{a}nek}
\affiliation{Computer Graphics Group, KSVI, Faculty of Mathematics and Physics, Charles University, Malostransk\'{e} n\'{a}m\v{e}st\'{i} 25, Prague 1 CZ-11800, Czechia}
\author{Jan Valenta}
\affiliation{Department of Chemical Physics and Optics, Faculty of Mathematics and Physics, Charles University, Ke Karlovu 3, Prague 2 CZ-121 16, Czechia}
\author{Yvo D. Mudde}
\affiliation{University of Amsterdam, Institute of Physics, Science Park 904, 1098 XH Amsterdam, The Netherlands}
\author{Peter Schall}
\affiliation{University of Amsterdam, Institute of Physics, Science Park 904, 1098 XH Amsterdam, The Netherlands}
\author{Kate\v{r}ina Dohnalov\'{a}}
\affiliation{University of Amsterdam, Institute of Physics, Science Park 904, 1098 XH Amsterdam, The Netherlands}
\altaffiliation[Presently]{K. Newell}
\email{k.newell@uva.nl}
\date{\today}

\begin{abstract}
For the development and optimization of novel light emitting materials, such as fluorescent proteins, dyes and semiconductor quantum dots (QDs), a reliable and robust analysis of the emission efficiency is crucial. The emission efficiency is typically quantified by the photoluminescence quantum yield (QY), defined by the ratio of emitted to absorbed photons. We show that this methodology suffers from a flaw that leads to underestimated QY values, presenting as a 'parasitic absorption'. This effect has not been described and/or corrected for in literature and is present already under common experimental conditions, therefore, it is highly relevant for a number of published studies. To correct for this effect, we propose a  modification to the methodology and a correction procedure.
\end{abstract}

\maketitle

\section{Introduction}
\begin{figure*}[hhht]
\centering
\includegraphics[width=0.9\textwidth]{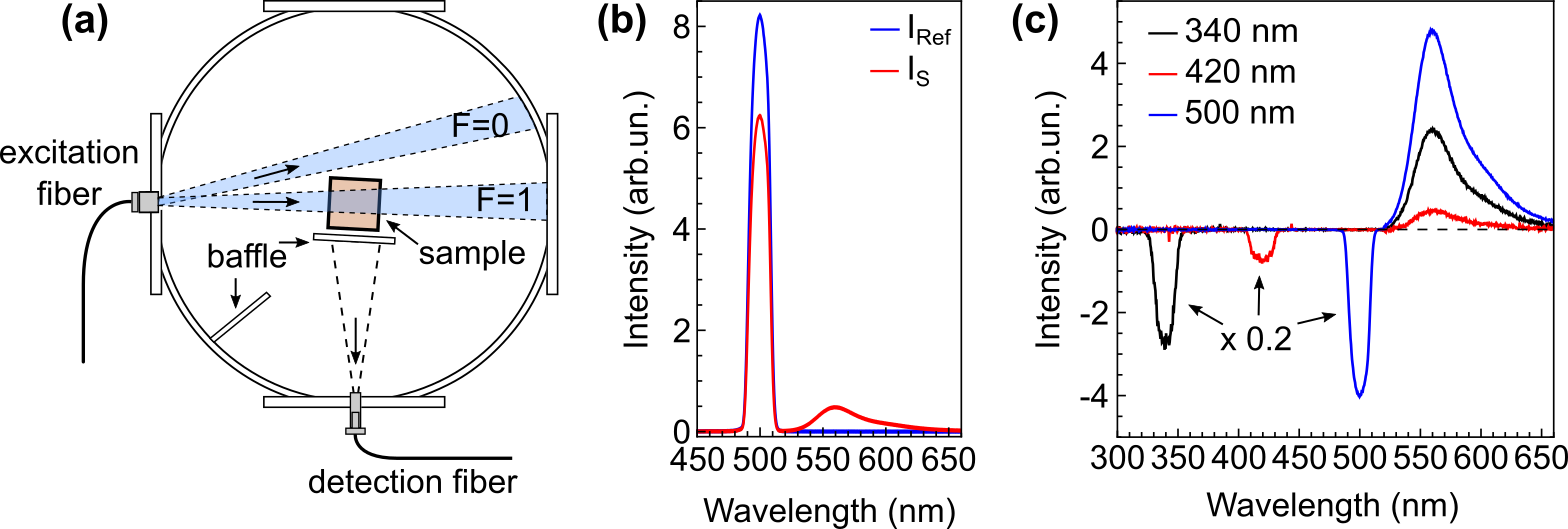}
\caption{(a) Schematic of the QY setup using the IS technique (top cross-section view): The sample in the center of the sphere is excited directly by a fraction F of the incoming excitation beam (direct excitation (F=1) and indirect excitation (F=0) are shown). Baffles prevent direct detection of the excitation and emission photons. See also Fig.~\ref{figureS1}. (b) Examples of typical spectra used for QY calculation recorded with the reference (blue) and sample (red), inserted inside the IS, for R6G in ethanol. (c) Three examples of spectra obtained by subtracting sample and reference measurements for R6G in ethanol for different excitation wavelengths.}
\label{figure1}
\end{figure*}

The development of novel materials for optoelectronics and photovoltaics requires an accurate and robust methodology for evaluation of the emission efficiency. This property is best quantified in terms of the absolute quantum yield (QY), defined as the ratio of emitted to absorbed photons. This methodology suffers from a critical flaw, which is for a great part independent of the particular experimental or instrumental implementation.  While many general guidelines exist for QY determination \cite{Wurth2015,Crosby1971,Wurth2013,Valenta2014,Wurth2011}, discussing e.g. the effects of re-absorption \cite{Ahn2007} and excitation geometry \cite{Faulkner2012,Wurth2015a}, the artifact we describe has not been identified until now, creating great urgency for a remedy. There is a variety of experimental techniques to estimate the QY \cite{Wurth2015,Crosby1971}, some relying on the comparison with a calibration standard of similar absorption and emission as the sample of interest, as well as a precisely known QY. Another more direct and accurate method implements an integrating sphere (IS), first demonstrated in 1995 by Greenham et al. \cite{Greenham1995} and later developed into a 3-step\cite{DeMello1997} and 2-step\cite{Mangolini2006} measurement, allowing the direct determination of the absolute number of emitted and absorbed photons by comparing the calibrated emission and absorption spectra of the studied sample to a suitable blank (Figure~\ref{figure1}a). This optical method is considered robust and reliable\cite{Wurth2015,Wurth2013} and has been standardized for the LED and display industry. Commercial QY devices based on the IS method can be purchased from several spectroscopic companies\cite{Horiba,Hamamatsu,Wurth2010}.\\

We demonstrate the artifact on the IS technique and show that it manifests itself in the underestimation of the QY, the magnitude of which depends on the absorption of the studied material. Hence, our findings have consequences not only for scientific studies where the QY is used to characterize and optimize new types of luminophors, but also in which the QY is used to uncover new photophysics, for example in quantum dots (QDs). Here, special care must be taken when comparing the QY of samples with diverse absorption as a result of different excitation energies\cite{Timmerman2011,Greben2015}, various QD sizes \cite{Mastronardi2012,Miller2012,Sun2015} or QD densities \cite{Greben2015} as interpretations of the observed QY dependencies could be affected considerably by the artifact. By comparison with simulations, we identify the artifact and provide and test a calibration procedure to correct for it. \\

\section{Results}
We measure the QY using the standardized two-step IS measurement (Figure~\ref{figure1}a and \ref{figureS1}) \cite{Mangolini2006}, where the emission and transmission spectra of the sample ($S$) are compared to that of a blank reference sample ($Ref$). The absolute numbers of emitted ($N_{em}$) and absorbed ($N_{abs}$) photons are obtained by integrating over the excitation ($exc$) and emission ($em$) spectral peaks respectively:
\begin{align}
QY=\frac{N_{em}}{N_{abs}}&=\frac{N_{S}^*-N_{Ref}^*}{N_{Ref}-N_{S}}=\nonumber\\
&=\frac{\int^{em}[I_S(\lambda)-I_{Ref}(\lambda)]C(\lambda)d\lambda}{\int^{exc}[I_{Ref}(\lambda)-I_{S}(\lambda)]C(\lambda)d\lambda} \label{eq1}
\end{align}

Here $N$ and $N^*$ denote the integrated intensities at the excitation and emission wavelength respectively, $I(\lambda)$  the detected spectrum and $C$ the correction factor for the spectral response of the detection system. Using this setup, we measure the QY of three different light emitting materials: an organic dye and two types of inorganic QDs, all in a wide range of excitation wavelengths and for various material concentrations as commonly encountered in the literature. \\

First we study Rhodamine 6G (R6G) in ethanol, which is a well-known fluorescent dye that is commonly used as a calibration standard for comparative QY measurements and has a high reported QY of about 95\%\cite{Kubin1982}. We prepared several solutions of R6G in ethanol, with concentrations between ~120 and 6 $\mu$M. Within this range, the spectral shape of the absorption coefficient is unaltered (Fig.~\ref{figureS2}a), indicating the absence of clustering effects\cite{Wurth2012} or other material changes. Also, the photoluminescence (PL) lifetime in the studied range is independent of excitation wavelength and concentration (Fig.~\ref{figureS2}c), showing that the internal emission efficiency is constant within the studied range. To evaluate the QY, we compare the emission and transmission spectra of the sample (R6G in ethanol) to that of a reference (ethanol) as shown in Fig.~\ref{figure1}b. From the difference, we obtained the total number of emitted and absorbed photons (Fig.~\ref{figure1}c). \\

\begin{figure*}[ht]
\centering
\includegraphics[width=0.9\textwidth]{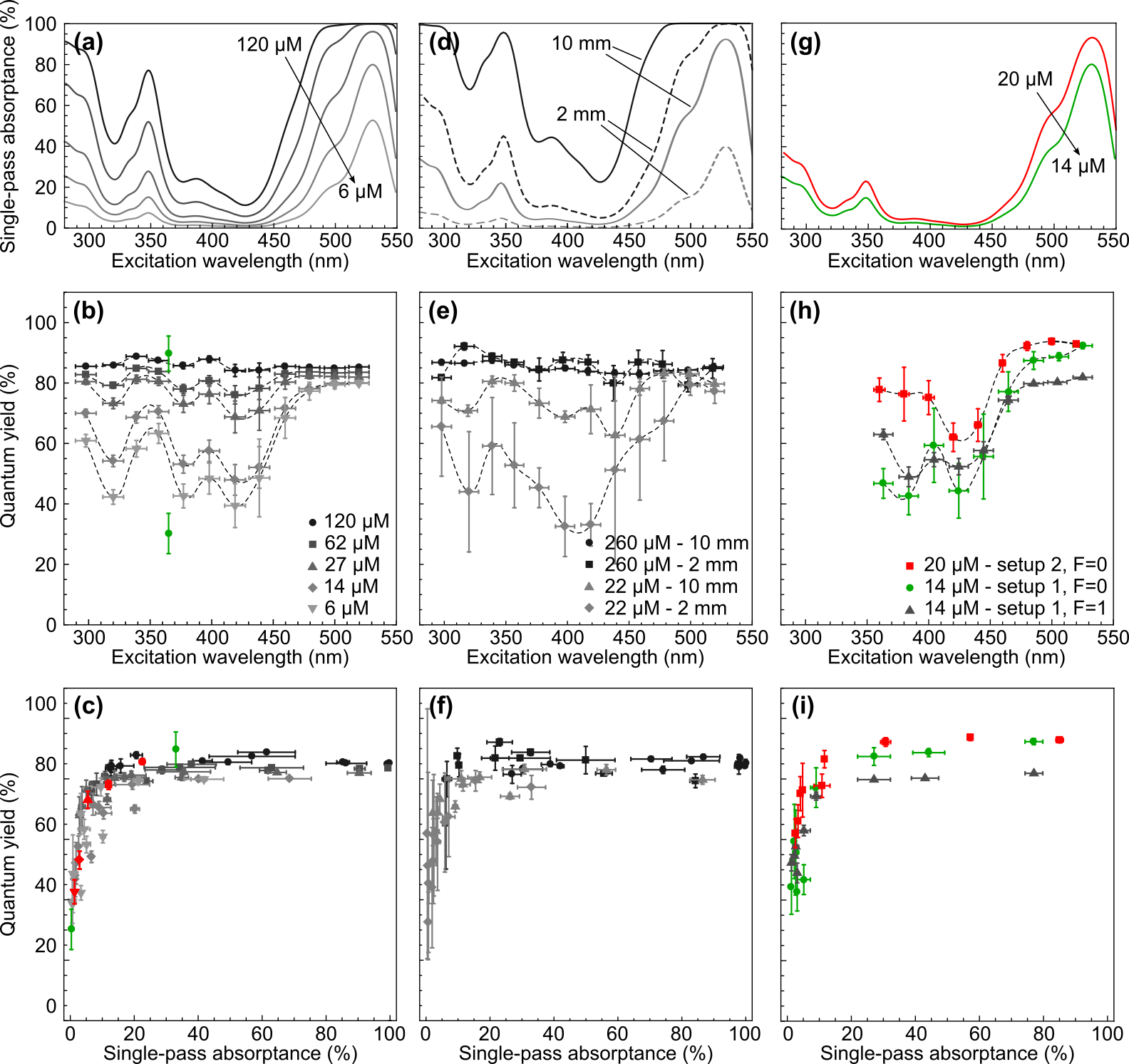}
\caption{The QY of R6G in ethanol for: (a-c) Varying  excitation wavelength and concentration , compared to literature values, (d-f) different thickness of the cuvette, i.e. optical path length and (g-i) direct and indirect excitation conditions and measured in two different setups (described in Material and Methods). Legend in the middle panels is shared also for the above and below panels. The single-pass absorptances are shown in (a,d,g), the experimentally determined QYs as a function of excitation wavelength are shown in (b,e,h) and the QY data from (b,e,h) replotted as a function of the single-pass absorptance are shown in (c,f,i). All data are corrected for reabsorption using the procedure described by Ahn et al. \cite{Ahn2007}. For comparison with literature, the QY values as determined by Faulkner et al. \cite{Faulkner2012} for a 100 $\mu$M (the higher QY value) and 1 $\mu$M (the lower QY value) solution of R6G in ethanol are plotted in (b, c) in green.}
\label{figure2}
\end{figure*}

QYs obtained at various R6G concentrations excited between 300 and 520 nm are shown in Fig.~\ref{figure2}b. For the highest concentration of 120 $\mu$M, we find a QY of $\sim$86\% (note that the difference from the expected value of 95\% is discussed later and in Fig.~\ref{figureS3} of the Supplementary Information) that is roughly constant over the whole spectral range, which is expected due to the Kasha-Vavilov rule \cite{Wawilow1927}. As we decrease the sample’s concentration and hence also the sample’s absorption, the determined QY drops significantly, at some points to as low as 38\%. Interestingly, the same decrease in QY is observed when the concentration is fixed and the excitation wavelength is lowered below $\sim$460~nm, for which the absorption coefficient of R6G decreases significantly. From Fig.~\ref{figure2}b, it is obvious that the QY as a function of excitation wavelength for the lower concentrations roughly follows the absorption spectrum of R6G, depicted for comparison in Fig.~\ref{figure2}a. To investigate this relation, we plot the QY against the single-pass absorptance of the sample in Fig.~\ref{figure2}c and we find that the QY decreases significantly at low absorptance, independently whether the absorptance is lowered via the sample concentration or via the excitation wavelength. At a critical absorptance value $A_{crit}$ of 10-15\%, there is an abrupt change in the QY behavior, where for $A > A_{crit}$ the QY is constant and close to the expected literature value, but decreases continuously for $A < A_{crit}$. The same effect can be observed when both, the concentration and excitation wavelength, are fixed and the single-pass absorptance of the R6G solution is decreased by changing the optical path length through the sample, by using a thinner cuvette, as shown in Figs.~\ref{figure2}d-f. This shows that the effect is not intrinsic to the studied material and depends only on the sample’s absorption. We carefully confirmed this effect also under different experimental conditions and in a different experimental setup, as shown in Figs.~\ref{figure2}g-i and \ref{figureS4}. \\

\begin{figure*}[ht]
\centering
\includegraphics[width=\textwidth]{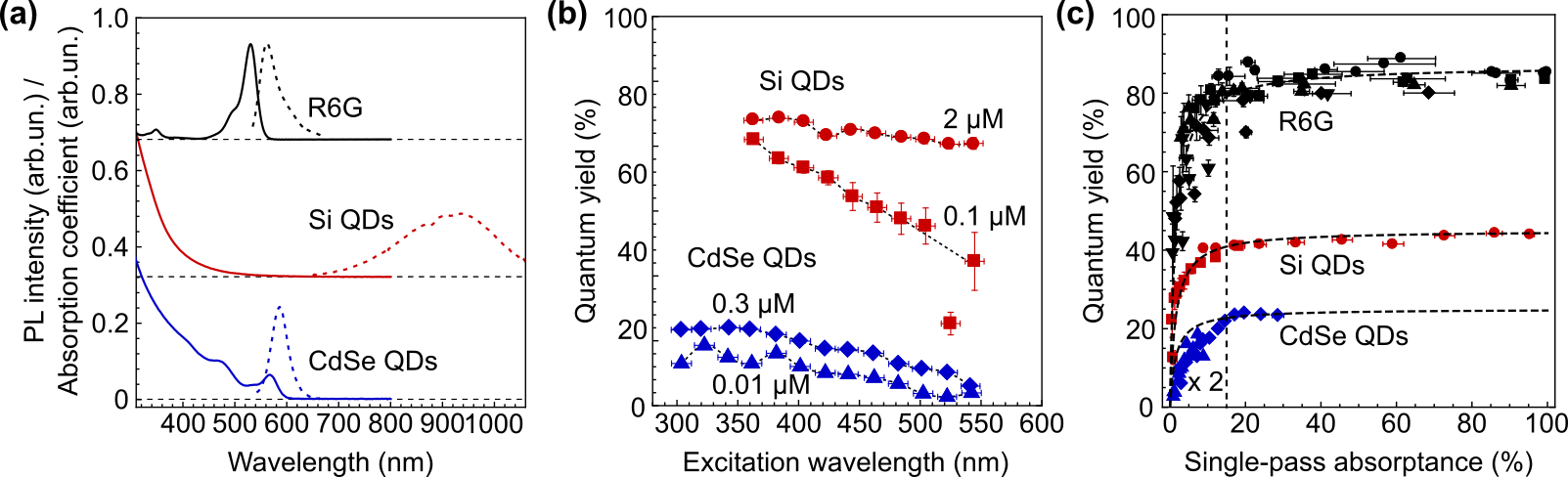}
\caption{(a) Absorption coefficients and PL spectra (dashed lines) for three types of samples: R6G in ethanol (black), Si QDs in toluene (red) and CdSe QDs in hexane (blue). (b) QY dependence on the excitation wavelength of CdSe QDs and Si QDs. The QY of the Si QDs below 360 nm is omitted since the solvent is non-transparent in this spectral region. The dashed lines serve as guides to the eye and the arrows point in the direction of the lower concentration samples. (c) QY versus single-pass absorptance. The data for R6G from Figure~\ref{figure2}c is shown for comparison (black symbols). The dashed lines serve as guides to the eye to emphasize the observed trend of the QY. The critical absorptance value $A_{crit}$ is indicated by the vertical dashed line.}
\label{figure3}
\end{figure*}

In addition to the R6G, we test the effect also on two suspensions of semiconductor QDs - Silicon QDs (Si QDs) and cadmium selenide based QDs (CdSe QDs). Both materials have very different optical properties from each other and from those of R6G, such as emission efficiency, absorption and emission spectra (Fig.~\ref{figure3}a). The most notable difference is the broad featureless absorption spectrum, resulting from the band-like dispersion in semiconductors. Also, Si QDs have a larger Stokes shift and a less abrupt absorption onset than CdSe QDs due to the indirect band-structure.  Despite these differences, we obtain a very similar effect: The QY of both types of QD materials decreases with longer excitation wavelengths (Fig.~\ref{figure3}b) below $A_{crit}$, similarly to the observations in R6G. Again, the effect is more pronounced for lower concentration. In the case of semiconductor QDs, one might argue that the process of emission and absorption is more complex than that of the organic dyes, and that therefore the validity of the Kasha-Vavilov rule might be weaker or not hold at all \cite{Timmerman2011,Valenta2014a}. With that line of reasoning, the observation of explicit QY dependence on various parameters could be interpreted in terms of (i) novel effects \cite{Timmerman2011, Greben2015,Mastronardi2012,Miller2012,Sun2015,Saeed2014,Chung2017}, (ii) as a result of size-polydispersity of the QD ensembles, leading to a broadening of the emission spectrum \cite{Miura2006}, which in turn could lead to excitation wavelength dependent QY via excitation of different subsets of the QD ensemble; or (iii) one could argue that the concentration of the QD dispersions is expected to affect QD interactions \cite{Timmerman2011,Limpens2015}, which strongly depend on the inter-particle distance. Nonetheless, whether the above interpretations hold in our case is questionable, since when plotting the QY of the QDs versus their single-pass absorptance (Fig.~\ref{figure3}c), the resemblance with the trend observed for R6G in Figs.~\ref{figure2}c,f,i is striking in terms of the shape of the QY-dependence and the similar value of $A_{crit}$, suggesting that the excitation- and concentration-dependence is for the largest part described by the same effect as observed for R6G. \\

\begin{figure}[ht]
\centering
\includegraphics[width=0.4\textwidth]{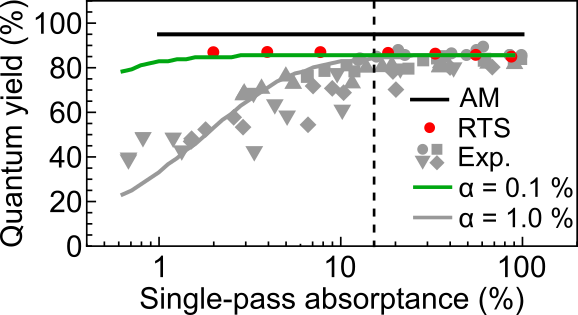}
\caption{Comparison of the experimental dependence of the QY of R6G on the single-pass absorptance (gray symbols, the same data as in Fig.~\ref{figure2}b) with simulated values by the AM (black line) and RTS (red symbols) approaches. For comparison is shown also peak value of the  QY probability distribution, obtained from the analytical expression Eq.~\ref{eq2} in the presence of measurement uncertainty in number of photons of $\alpha$=0.1\% (green) and 1.0\% (gray line).}
\label{figure4}
\end{figure}

From these experiments, where we observe the same decrease of QY when the single-pass absorptance of the sample is lowered either by lowering the sample's concentration, shifting the excitation wavelength or by shortening the optical path, and this independently of the studied material or experimental setup, we conclude that it is an artifact of the QY methodology. This flaw becomes even more apparent, when we compare our measurements to the simulations of the QY experiment in the IS setup. For this we use an analytical model (AM), that describes the light inside the IS by uniformly distributed photon fluxes in a uniform IS geometry (Fig.~\ref{figureS7})\cite{vanDam2018}, an approach which is a commonly applied in literature\cite{DeMello1997,Mangolini2006,Valenta2014,Faulkner2012}. In this approach, the measured intensities $I^*$ and $I$ in Eq.~\ref{eq1} are expressed in terms of the probability that emission and excitation light reaches the detector port via multiple reflections inside the IS (Fig.~\ref{figureS7}). We further validate this analytic model by numerical ray-tracing simulations (RTS), in which different photon paths are individually considered, thus yielding solutions even for non-uniform photon distributions and IS geometries, which allows for more exact modeling of the setup (Figs.~\ref{figureS1},\ref{figureS5} and \ref{figureS6}). The qualitative comparison between the simulations, analytical model and the experimentally obtained QYs is shown in Fig.~\ref{figure4}. Both, the AM and RTS approach, yield the expected flat dependence associated with a constant QY. Specifically, the AM data preserves the input value of the material’s QY of 95\%, as expected for R6G\cite{Kubin1982}, while the RTS approach yields a slightly lower QY of $\sim$87\%, in good agreement with our experimentally observed values for R6G with absorption above 15\%, showing the strength of the RTS approach for modelling of the specific IS setup. For more details for both approaches, AM and RTS, we refer to the Supplementary Materials and Ref.~\cite{vanDam2018}. Nevertheless, it is obvious that for lower absorptances, the experimental data strongly deviate from the predictions given by both, the AM and RTS simulation approaches (Fig.~\ref{figure4}).\\

First, we would like to show that part of this artifact originates from a statistical nature of the experiment, while AM and RTS approaches assumed a single (mean) value of number of absorbed and emitted photons $N_{abs}$ and $N_{em}$, neglecting their statistical distribution. Through the QY definition by Eq.~\ref{eq1}, these distributions give rise to a distribution of QY values, which as we will see in the following will no longer be centered around the value given by the means of $N_S$ and $N_{Ref}$. To account for the distribution of photons in real experiment, we assume the number of emitted photons $N_{em}$ and number of photons transmitted through reference and through sample $N_{Ref}$ and $N_S$, to be independent and normally distributed around a mean value $\mu$ with variance $\sigma^2$, i.e. $N\sim\mathcal{N}(x,\mu,\sigma^2)$. We note that for very low photon counts (below $N\sim$100), one would need to consider Poisson distribution instead, which, however, will lead to the same output, as we have shown in Ref.~\cite{vanDam2018}. To derive distribution of QY values, we first derive the probability distribution for number of emitted and absorbed photons. The number of emitted photons is given by $N_S^*\sim\mathcal{N}(x,\mu_S^*,\sigma_{S*}^2)$, as we can neglect emission from the reference $N_{Ref}^*$, which is typically zero or very small. The number of absorbed photons is given by the difference distribution of the $N_{Ref}$ and $N_S$, i.e. $N_{abs}\sim\mathcal{N}(x,\mu_{Ref}-\mu_S,\sigma_{Ref}^2+\sigma_S^2)$. The probability distribution of the QY values is then given by the ratio of the distributions of $N_{em}$ and $N_{abs}$. By explicitly computing this ratio of distributions, it can be shown that the resulting probability distribution of QY values is \cite{blejec}:

\begin{align}
p_{QY}(x)=&\frac{\theta}{\pi(x^2+\theta^2)} \nonumber  \\
\times &[\sqrt{2\pi}B(x)\Phi(B(x))e^{-\frac{C(x)}{2}}+K]
\label{eq2},
\end{align}
where 
\begin{align*}
B(x)=\frac{\alpha_{abs}x+\alpha_{em}\theta}{\alpha_{em}\alpha_{abs}\sqrt{x^2+\theta^2}}\\
\Phi(x)=\int_{-\infty}^x \frac{1}{\sqrt{2\pi}}e^{-\frac{1}{2}u^2}du\\
C(x)=\frac{(\alpha_{abs}\theta-\alpha_{em}x)^2}{\alpha_{em}^2\alpha_{abs}^2(x^2+\theta^2)}\\
K=exp(-\frac{\alpha_{em}^2+\alpha_{abs}^2}{2\alpha_{em}^2\alpha_{abs}^2}).
\end{align*}

With parameter $\theta=\frac{\sigma_{em}}{\sigma_{abs}}$ and relative measurement uncertainties $\alpha_i=\frac{\sigma_i}{\mu_i}$, with $i=abs, em$. This distribution is a Cauchy-like distribution (discussed in greater detail in Ref.~\cite{vanDam2018}) - a non-trivial function with complex shape, strongly depending on the divisor distribution $N_{abs}$. We plot the position of the resulting most-likely QY value (the position of the peak of the positive part of the ratio probability distribution from Eq.~\ref{eq2}) for two distinct values of the measurement uncertainty $\alpha$ in Fig.~\ref{figure4}: For a very precise measurement with $\alpha$ = 0.1\% (green curve), the resulting most-likely value of QY remains flat while for a larger spread in values represented by $\alpha$ of e.g.  1.0\% (gray curve), we obtain underestimated QY below certain critical absorptance, which qualitatively agrees with the measured data. However, experimentally determined $\alpha$ is 0.3\%, which means that the statistical nature of the experiment can only explain part of the QY underestimation.\\

\begin{figure}[ht]
\centering
\includegraphics[width=0.4\textwidth]{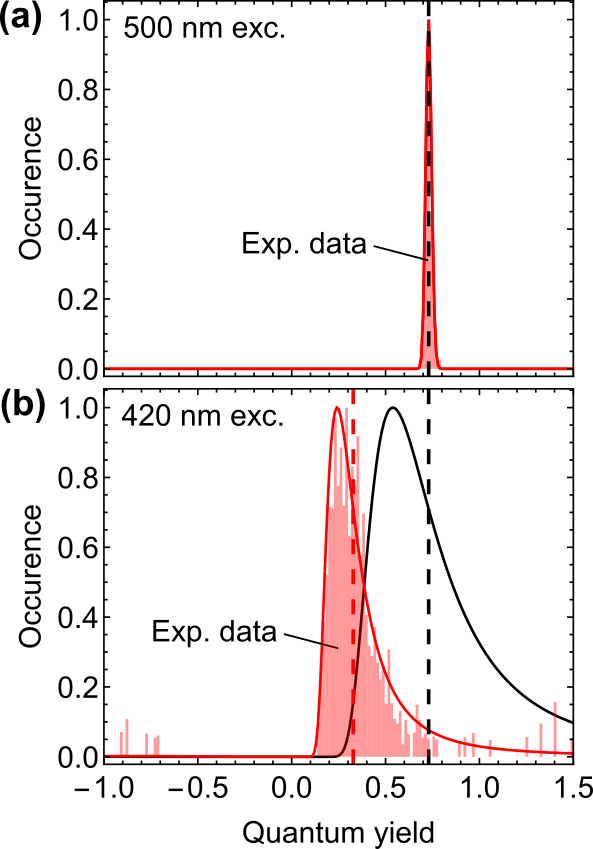}
\caption{Measured QY distribution of 6 mm solution of R6G in ethanol under (a) 500 and (b) 420 nm wavelength excitation. Solid lines represent fits using the ratio distribution (Eq.~\ref{eq2}) assuming an absorptance offset of 1.7\% (red) and without absorption offset (black).}
\label{figure5}
\end{figure}

This is further illustrated by the measured QY distribution function, obtained by repeating the QY measurement 36 times. The resulting QY histograms for R6G in ethanol are shown in Fig.~\ref{figure5}. For the high absorption at 500 nm excitation (Fig.~\ref{figure5}a) the measured QY histogram is narrow and centered around the expected value (black dotted line). For the low absorption at 420 nm excitation (Fig.~\ref{figure5}b), we obtain a broad and skewed histogram with most-likely QY value (peak of the red histogram) far from the expected value (black dotted line). Using our analytical model (AM) with normally distributed photon counts with distribution parameters experimentally evaluated directly from the histogram of $N_{em}$ and $N_{abs}$, we fit the QY data using the Eq.~\ref{eq2} (black line). Clearly, the curve is shifted and does not properly fit the QY histogram. This might appear surprising, since the experimental QY data in Fig.~\ref{figure4} (gray symbols) were fitted relatively well by the Eq.~\ref{eq2} using experimental uncertainty of $\alpha$=1.0\%(gray line). However, as we already mentioned, the measured experimental uncertainty is only $\alpha$=0.3\%, hence the discrepancy. Therefore we conclude, that the skewed QY distribution caused by the statistical nature of the experiment cannot on its own explain the QY underestimation, which is observed to be nearly twice as higher.\\

\begin{figure}[ht]
\centering
\includegraphics[width=0.4\textwidth]{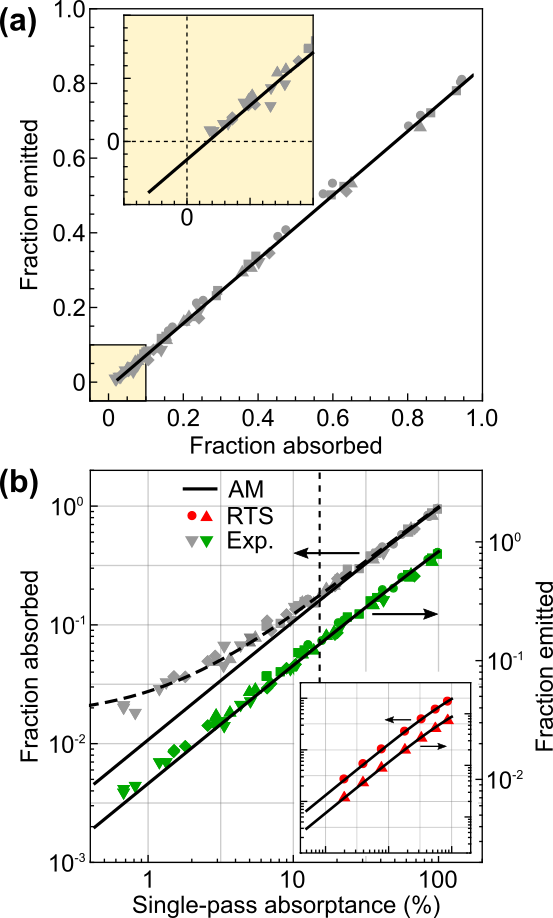}
\caption{(a) The normalized number of emitted photons against the normalized number of absorbed photons obtained from multiple measurements on R6G in ethanol. The black line is a linear fit yielding a QY (slope) of 86\%. The line does not intersect at the origin however, but at an absorptance of 1.7\% (zoom in the inset). (b) Experimentally determined dependences of the absorbed (gray symbols, left axis) and emitted (green symbols, right axis) photon fractions on the single-pass absorptance of the sample. Black line indicates the simulated dependence using the noiseless AM approach. Dotted line fits the experimental data with absorption offset of 1.7\%. In the inset is shown comparison of the AM approach (black lines) with the RTS approach (red symbols - fraction of absorbed (circles) and emitted (triangles) photons). Both approaches are in excellent agreement. }
\label{figure6}
\end{figure}

To better understand this mismatch, we separately look at the number of absorbed and emitted photons. In Fig.~\ref{figure6}a we plot the $N_{em}$ as a function of $N_{abs}$, which follows a linear dependence with a slope 0.86, as expected for QY of 86\%. We find that this linear dependence does not cross axes at the origin (inset in Fig.~\ref{figure6}a). This shift can be interpreted as either an emission underestimation, or an absorption overestimation. To determine which is the case, we plot  $N_{em}$ and $N_{abs}$ as a function of the single-pass absorptance in Fig.~\ref{figure6}b. We find that $N_{em}$ decreases roughly linearly with the single-pass absorptance, but $N_{abs}$ deviates from the linear dependence for below $A\sim$15\%, value coinciding with the critical absorption $A_{crit}$ below which we observe the underestimated QY (dashed vertical line). This is again compared with the noiseless AM approach (black lines), where both $N_{em}$ and $N_{abs}$ decrease linearly, in agreement with RTS simulations shown in the inset. We conclude that the issues with the underestimation of the QY originate from an overestimation of $N_{abs}$, while quite contra intuitively, $N_{em}$ and its associated uncertainty has only negligible effect.\\ 

The shift in $N_{abs}$ evaluated from both dependencies in Fig.~\ref{figure6}a and b is $\sim$1.7\%. We note that this shift is of unclear origin and might be setup-dependent (in comparison, we obtained $\sim$1\% shift on QY setup 2). Taking this shift into account in our model, we fit the QY histogram in Fig.~\ref{figure5}b again using Eq.~\ref{eq2} and obtain excellent agreement (red curve) with respect to the fit's position and shape. Therefore, we conclude that by considering the combination of both contributions, the skewness of the  QY distribution (Eq.~\ref{eq2}) and the shift in $N_{abs}$ (Fig.~\ref{figure6}), we can describe very precisely the experimentally measured QY values in low and high absorbing materials (Fig.~\ref{figure5}), which can be used for the remedy. \\

\section{Correction procedure}

While the identified issues critically affects QY values for samples with lower absorption, we can correct for it as follows. First, we determine the critical absorption value $A_{crit}$ for each specific QY setup using a reference material - a fresh, high quality QY standard, such as R6G. Please note that it is essential to take a special care for re-absorption, possible lifetime changes, aggregation, etc. (see our detailed discussion above and in Supplementary Materials). Next, since the $A_{crit}$ can be expected in range 5-20\% in a typical QY setup, one needs to measure QY dependence on the single-pass absorptance in a wide range of absorptances, set at least between 1 and 30\%. Such a broad absorption range can be reached either by varying (i) the excitation wavelength, (ii) the sample’s concentration and/or (iii) the sample’s optical thickness, or by a combination of the above, since the artifact manifests itself independently of how the absorption is changed. $A_{crit}$ is given by the single-pass absorptance at which the measured QY deviates from the expected constant value, as shown in Figs.~\ref{figure2}c,f,i. Finally, the QY of the sample of interest ($QY_S^{meas}$) is measured, from which a more reliable QY estimate ($QY_S^{corr}$) can be obtained in the range $A < A_{crit}$, by comparison of the obtained QY dependence of the calibration standard $QY_{Cal}(A)$, with the unbiased value $QY_{Cal}(A>A_{crit})$:

\begin{eqnarray}
QY_S^{corr}(A)=QY_S^{meas}(A)\frac{QY_{Cal}(A>A_{crit})}{QY_{Cal}(A)}
\label{eq3},
\end{eqnarray}

\begin{figure*}[ht]
\centering
\includegraphics[width=0.85\textwidth]{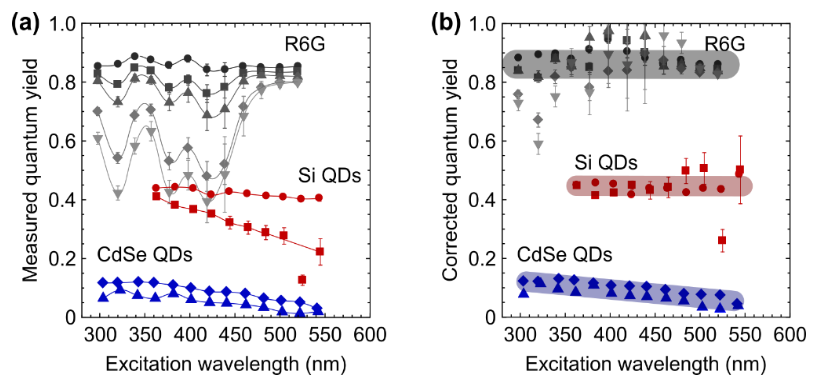}
\caption{QY versus excitation wavelength, before (a) and after (b) correction for the absorption dependence of the QY methodology using Eq.~\ref{eq3}. Different symbols represent different sample concentrations, as in Figs.~\ref{figure2} and \ref{figure3}. After correction, the QY of the Si QDs is constant at $\sim$45\% and independent of the sample concentration. Part of the excitation and concentration dependence of the QY of the CdSe QDs persists after correction, potentially related to the ligands on the surface\cite{Greben2015,Grabolle2009,DeRoo2016,Semonin2010}. The correction function was obtained by taking a moving average of the QY versus single-pass absorption dependence of R6G in Fig.~\ref{figure2}c. For reference, the corrected QY for the R6G is also shown. }
\label{figure7}
\end{figure*}

We applied this correction procedure to the QY measurements, presented in Figs.~\ref{figure2} and \ref{figure3}, with a result shown in Fig.~\ref{figure7}. Indeed, the vast majority of the observed spectral and concentration dependence of the QY can be explained by the described artifact. Only for this particular sample of CdSe QDs, part of the QY decrease remains after correction, which points towards some inherent physical QY effect, not linked to the artifact. Interestingly, this is also accompanied by a nonlinear dependence of the emitted photon fraction $N_{em}$, which was not observed in R6G or Si QDs (Fig.~\ref{figureS8}). This indicates that as an additional, independent test, one can plot fraction of emitted photons $N_{em}$ as a function of the single pass absorptance of one's sample for a broader range of absorptances (on the lower side) to find whether the lower QY is purely due to the described artifact, or has some interesting physical meaning. Our proposed procedure corrects for the artifact, and therefore ensures reliable QY estimates in a much broader range of experimental conditions and sample variations. This procedure can be applied to any setup and we strongly suggest this to be done also for any commercial IS setups and relative and comparative QY measurements. \\

\section{Conclusion}
In conclusion, we report on a critical artifact generally present in the QY methodology, leading to underestimation of the QY value under common experimental conditions.  This artifact is partly caused by skewed QY distribution resulting from statistical nature of the experiment \cite{vanDam2018} and partly due to a shift in $N_{abs}$ of unclear origin. Most importantly, this artifact manifests itself independently of the type of material and of the specific geometry of the experimental setup. We prove that this artifact strongly depends on material's absorption and hence is critical not only for studies where QY is evaluated to as an important material characterization, but also in studies where materials with different absorption are compared, or a single material is studied in a broad range of excitation wavelengths, concentrations or other parameters that affect the material’s absorption. Even for the calibration standard R6G, both non-resonant excitation and insufficient concentration lead to a strongly underestimated QY (Fig.~\ref{figure2}). The QY is frequently used also to characterize novel materials such as semiconductor QDs, to study the emission efficiency dependence, e.g., on size\cite{Mastronardi2012,Sun2015} and density \cite{Miller2012} of the QDs, excitation energy \cite{Timmerman2011,Valenta2014a} or to show ligand instability \cite{Greben2015}, etc.. A critical re-assessment of these results and of the methodology is therefore needed, since the size, density and excitation energy are intimately linked to the magnitude of the single-pass absorption of the studied samples. Our proposed calibration procedure corrects for this intrinsic absorption-dependence and eliminates the artifact. In this way it possible to establish robust and reliable QY measurements for materials developed for e.g. bio-imaging, bio-sensing or optoelectronic devices.

\section*{Materials and Methods}
\textbf{Sample preparation:} Rhodamine 6G (R6G) from Sigma-Aldrich was dissolved in UV-grade ethanol (Merck KGaA, Uvasol), from which different concentrations were prepared by dilution. The concentration was estimated by comparison of the measured absorption coefficients with the value specified by Birge in Ref.~\cite{Birge1987}. For all optical measurements, $\sim$1.5 mL of solution is contained in a UV spectroscopy-grade quartz cuvette (Hellma Analytics, 111-QS). CdSe/ZnSe/ZnS core/shell/shell QDs (CdSe QDs) in hexane were purchased from Center for Applied Nanotechnology (CAN) GmbH (CANdots Series A CSS). High efficiency Si QDs were obtained from the group of Prof. J.G.C. Veinot (Department of chemistry, University of Alberta), prepared following the synthesis described in Refs.~\cite{Kelly2010,Hessel2006}. The concentration was estimated from the measured absorption coefficient and the absorption cross-sections reported in Ref.~\cite{Valenta2016}.\\

\textbf{Single-pass absorptance:} Transmittance of the sample ($T_S$) and reference ($T_{Ref}$) was measured using a dual-beam spectrophotometer (Perkin Elmer, Lambda 950), from which the sample’s single-pass absorptance was evaluated: $A=1-T_S/T_{Ref}$. For this, it is assumed that the reflectance of the sample and reference is the same. \\

\textbf{Absolute quantum yield:} \\
\textit{Setup 1}: For determination of the QY we use a standard integrating sphere (IS) setup, shown schematically in Figs.~\ref{figure1}a and \ref{figureS1}. To illuminate our sample, we use a stabilized Xenon lamp (Hamamatsu, L2273) coupled to a double-grating monochromator (Solar, MSA130). The excitation beam is split using a spectrally broad bifurcated fiber (Ocean Optics, BIF600-UV-VIS), where we use one part to monitor the fluctuations of the excitation intensity using a power-meter (Ophir Photonics, PD300-UV) and the other part to excite the sample. We use a collimator lens to reduce the spot-size at the sample position to enable direct excitation of the sample ($F$=1). The samples are suspended using an aluminium holder in the center of the IS (internal walls made of Spectralon$^{TM}$ (PTFE), Newport, 70672) with a diameter of 10 cm. The use of such type of holder is verified using ray-tracing simulations (see discussion on Fig.~\ref{figureS6}). Light is detected using a second spectrally broad optical fiber (Ocean Optics, QP1000-2-VIS-BX), coupled to a spectrometer (Solar, M266) equipped with a CCD (Hamamatsu, S7031-1108S). All measurements are corrected for the spectral response of the detection system, which we determine by illuminating the IS via the excitation port with a tungsten halogen calibration lamp (standard of spectral irradiance, Oriel, 63358) for the visible range, and a deuterium lamp (Oriel, 63945) for the UV range (< 400 nm). The measured calibration spectrum is corrected for the spectrometer’s stray-light contribution. QY is evaluated using Eq.~\ref{eq1}. Re-absorption effects were corrected for using the procedure described by Ahn et al.~\cite{Ahn2007} by comparing the measured PL spectrum with that of low concentration sample for which re-absorption is negligible (dashed line in Fig.~\ref{figureS2}b). Error estimates are obtained following Chung et al.~\cite{Chung2015}.\\

\textit{Setup 2}: The integrating sphere (IS) of 10 cm diameter has the inner surface covered by the Spectraflect$^{TM}$ coating. The excitation is provided by a laser-driven light source (LDLS, Energetiq, EQ-99X) coupled to the 15-cm monochromator (Acton SpectraPro SP-2150i). The monochromatized light (bandwidth of about 10 nm) is guided to the IS via a silica fiber bundle. The output signal from the IS is collected by another fused-silica fiber bundle placed in the direction perpendicular to the excitation axis and is shielded by three baffles against the direct visibility of both the excitation source and the sample. The end of the fiber bundle (which has a stripe-like shape) is imaged to the input slit of an imaging spectrograph (focal length of 30 cm) and a liquid-nitrogen-cooled CCD camera is used for detection. The spectral sensitivity of the complete apparatus is calibrated over the broad UV-NIR spectral range (300-1100 nm) using two radiation standards (Newport Oriel): a 45-W tungsten halogen lamp (above 400 nm) and a deuterium lamp (below 400 nm). Special attention is paid to avoiding stray-light effects in the spectrometer. 

\section*{Author contributions}
BvD has carried out the experimental measurements and analyzed the data. BvD, BB, GD and KD performed simulations with the analytical model. IK, AW and JK performed ray-tracing simulations. KD with YDM initiated this research and did preliminary experimental measurements. JV provided the independent experimental setup and analysis and initiated the ray-tracing approach. All authors have contributed to the discussion and final version of the manuscript.

\begin{acknowledgments}
The authors acknowledge Dutch STW funding (BvD, KD), FOM Projectruimte No. 15PR3230 (KD), MacGillavry Fellowship (KD); projects No. 16-22092S (JV), No. 16-08111S (AW) and No. 16-18964S (IK,JK) of the Czech Science Foundation; and support from the ESIF, EU Operational Programme Research, Development and Education, and from the Center of Advanced Aerospace Technology (CZ.02.1.01/0.0/0.0/16 019/0000826), Faculty of Mechanical Engineering, Czech Technical University in Prague (GD). The authors would like to the thank prof. T. Gregorkiewicz (University of Amsterdam) for facilitating this project, S. Regli and J. Veinot (University of Alberta, Canada) for providing the Si QDs, I. Sychugov (KTH-Royal Institute of Technology) for an independent control measurement and M. Hink (University of Amsterdam) for assistance with the time-resolved measurements. Finally, access to computing and storage facilities owned by parties and projects contributing to the National Grid Infrastructure MetaCentrum provided under the program "Projects of Large Research, Development, and Innovations Infrastructures" (CESNET LM2015042) is greatly appreciated.
\end{acknowledgments}

\beginsupplement
\section*{Supplementary materials}

\begin{figure*}[ht]
\centering
\includegraphics[width=\textwidth]{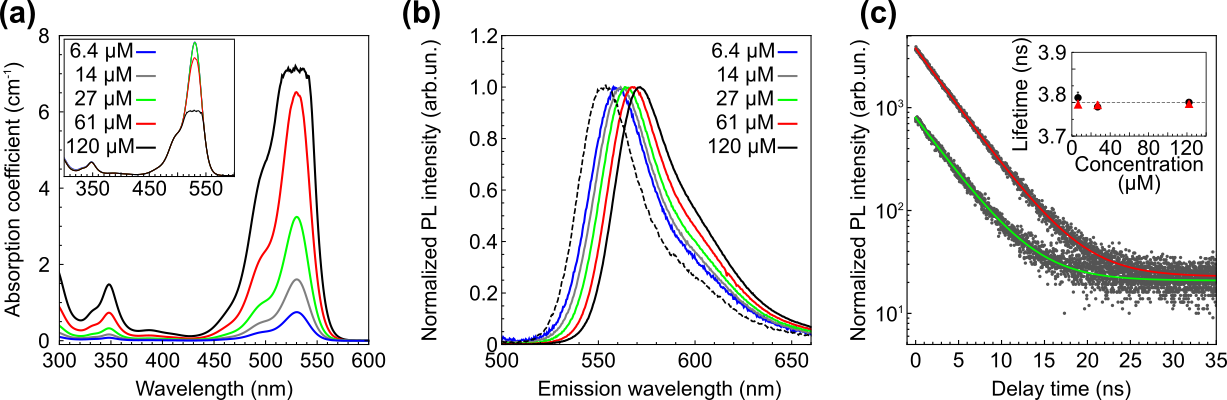}
\caption{\textbf{Rhodamine 6G.} (a) Absorption coefficient of different concentrations of R6G in ethanol normalized to the value. Inset shows the normalized absorption coefficient. The flattening of the absorption peak around 530 nm with higher concentration is due to absorption saturation effects, since transmission is close to zero \cite{Duyens1956}. (b) Normalized PL intensity of different concentrations of R6G in ethanol under 480 nm excitation. The dashed line shows the PL spectrum of a 1.6 $\mu$M concentration sample, measured outside of an IS, for which re-absorption effects are negligible. (c) Time-resolved PL intensity excited at 485 nm for a 120 $\mu$M (red) and 6.4 $\mu$M concentration of R6G in ethanol. Inset: lifetimes for different concentrations under 440 nm (black) and 485 nm (red) wavelength excitation. The lifetimes are obtained from a mono-exponential fit yielding $\sim$3.8 ns.  }
\label{figureS2}
\end{figure*}

\subsection*{Optical properties of Rhodamine 6G}
Absorption, photoluminescence (PL) spectra and decay are shown in Fig.~\ref{figureS2}. The time-resolved PL was measured using an inverted confocal microscope (Olympus, FV1000) equipped with a TCSPC module (Picoquant, Picoharp). For excitation, we used a pulsed 440 or 485 nm wavelength laser diode (Picoquant, LDH-P-C-440B or LDH-P-C-485) operated at 20 MHz (~2.7$\mu$W or 0.4$\mu$W) focused to diffraction limited spot by a 60x water immersion objective (Olympus, UPLS Apo, NA=1.2). Light is collected by the same objective lens, filtered through a 562/40 band-pass filter and detected using an avalanche photon detector (MPD, PDM). For the measurements the R6G solutions were contained in a 96-well plate. \\

\begin{figure*}[ht]
\centering
\includegraphics[width=0.9\textwidth]{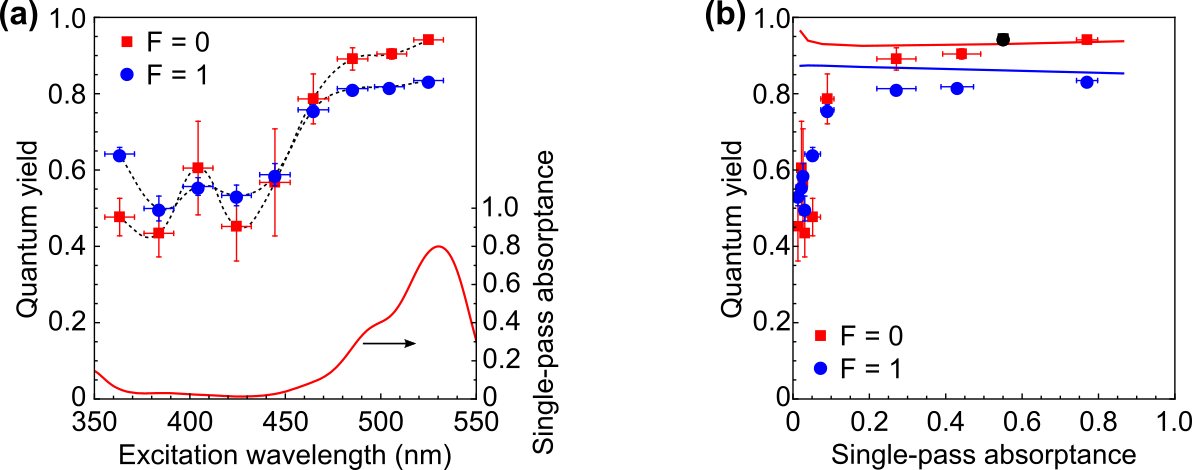}
\caption{(a) QY versus excitation wavelength of R6G in ethanol (~14 $\mu$M) with the sample under direct (blue) and indirect (red) illumination conditions. The solid line shows the single-pass absorption of the sample. (b) QY plotted against the single-pass absorption. The solid lines show the QY values simulated by RT under direct (blue) and indirect (red) excitation for a real QY of 95\%. For these simulations we assume an IS geometry similar to that used to obtain the experimental QY values, which includes aluminum parts to hold the optical fibers and cuvette. Black points show QY values obtained by RT simulations (circle, F=1; square, F=0) where we artificially replaced these aluminum parts by diffuse reflective parts. }
\label{figureS3}
\end{figure*}

The maximum QY that we find experimentally for R6G in Fig.~\ref{figure2} is $\sim$86\%, which is underestimated with respect to the literature value of 95\%~\cite{Kubin1982}. In contrast, under indirect excitation conditions we measure a QY which approaches 95\% as shown in Fig.~\ref{figureS3}. To understand this difference, we simulate the QY measurements under direct (F=1) and indirect (F=0) excitation conditions for the IS geometry as described in the materials section and shown in Figs.~\ref{figure1} and \ref{figureS1}. The ray tracing simulation (RTS) approach shows that in our specific geometry the QY is underestimated by a factor of $\sim$0.91 under direct excitation, whereas under indirect excitation a QY estimate close to the real value is obtained (Fig.~\ref{figureS3}b). These simulated values are in good agreement with the experimental QY values obtained for direct and indirect excitation (Fig.~\ref{figureS3}b). Most likely, the underestimation is related to the specific geometry of the IS setup 1, which contains home-built aluminum parts that serve as holders for the cuvette and the fibers. Indeed, when we replace the aluminum in our RTS approach with a Lambertian scatterer, both under direct and indirect excitation we obtain the correct QY value (black points in Fig.~\ref{figureS3}b). We note here that aluminium part are firmly excluded as a possible source of the artifact or absorption shift, evidenced by the RTS results. The ability to deal with non-uniform IS geometry shows the strength of RTS for modeling of the IS optics, which is not possible with the AM since it assumes a uniform IS geometry. \\

\begin{figure*}[ht]
\centering
\includegraphics[width=0.85\textwidth]{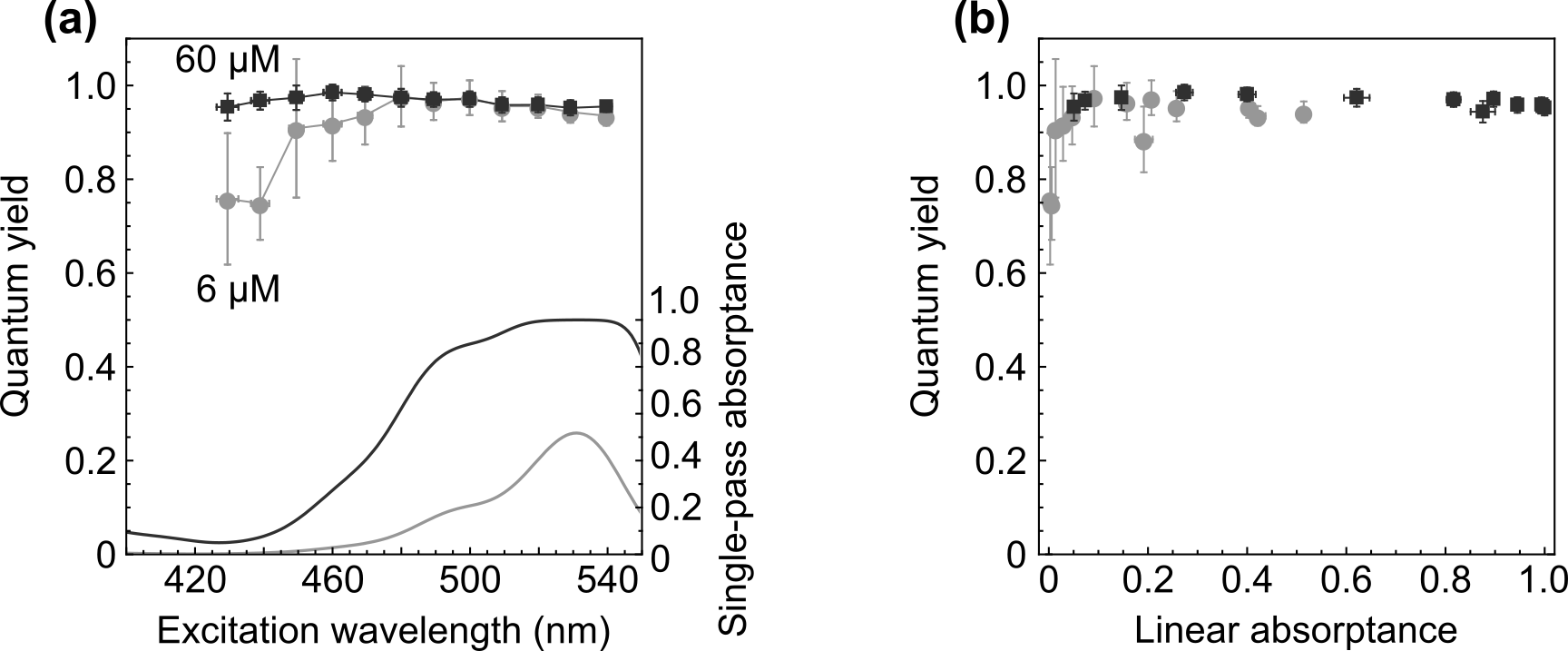}
\caption{QY of ~60$\mu$M (black) and ~6 $\mu$M (gray) solution of R6G in ethanol under direct excitation plotted against the excitation wavelength (a) and single-pass absorption (b). Measurements were carried out in the 'reference' QY setup 2.}
\label{figureS4}
\end{figure*}

Despite the fact that our geometry leads to slightly underestimated QY values, we note that our simulations (solid lines Fig.~\ref{figureS3}b) reveal no dependence of the QY on the absorption. Furthermore, measurements carried out in a completely different IS setup, shown in Fig.~\ref{figure2}g-i and \ref{figureS4}, again show an underestimation of the QY for weakly absorbing samples. Hence we conclude that the absorption-dependence of the QY does not arise from the specific geometry of our IS setup, but results from the QY methodology itself (as discussed in more detail in the main text).\\ 

\subsection*{Ray-tracing simulations}
\begin{figure*}[ht]
\centering
\includegraphics[width=\textwidth]{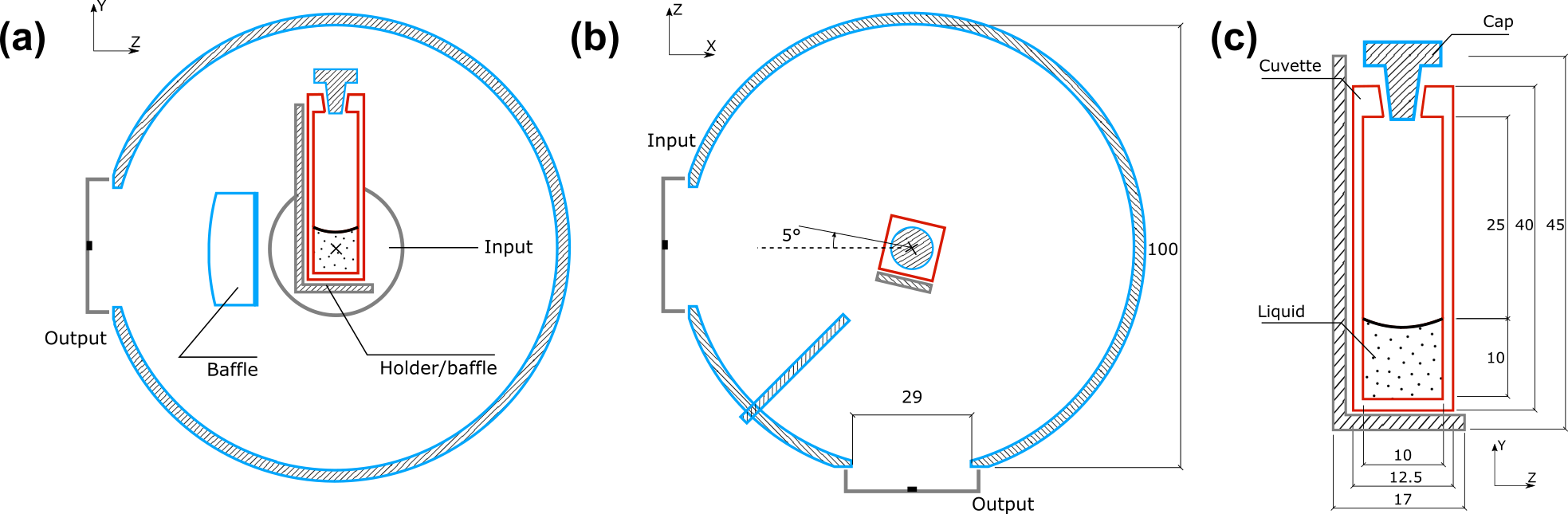}
\caption{\textbf{Schematic of the integrating sphere used for quantum yield measurements for setup 1.} Side view (a), top view (b) and zoom in of cuvette. (c) Different materials are indicated by colors; Lambertian diffuse surfaces with high reflectance (blue), quartz glass materials (red) and brushed aluminum surfaces (gray). Numbers indicate the dimensions in mm.}
\label{figureS1}
\end{figure*}

For the ray tracing simulation (RTS) we model the geometry of our experimental setup as shown in Fig.~\ref{figureS1}. The sphere is modeled as a well subdivided icosahedron with approximately $8.10^3$ triangles. The light distribution inside the IS is described by the radiative transfer equation \cite{Chandrasekhar1960} that can be solved by means of Monte-Carlo simulations. Multiple paradigms how to do such simulations exist \cite{Dutre2002,Veach1997,Pharr2010}; due to the combination of materials present in our setup, here we restrict our model to tracing light particles (“photons”) from the input port (contains the excitation source) to the output port (contains the sensor). By “photon” we denote a simulation particle with associated weight which represents the differential flux carried by the particle \cite{Veach1997}. For simplicity we model neither polarization, nor reabsorption events and we only keep notion of two types of photons characterized by two respective wavelengths: excitation and emission photons ($\lambda_{exc}$ and $\lambda_{em}$). Excitation photons colliding with the active part of the sample can be transformed into emission photons. \\

Photons interacting with any surface or absorbing volume are attenuated and scattered according to an appropriate phenomenological model: (i) The IS surface, baffle and the cuvette’s cap are modeled by a Lambertian model with reflectance at both wavelengths set to 0.97; (ii)  Scattering on glass and aluminum parts is modeled according to Fresnel equations for unpolarized light; (iii) Scattering/absorption in the volume of the active sample is modelled according to the radiative transfer equation where we set the absorption coefficient in the range 0.02 cm$^{-1}$ to 2 cm$^{-1}$ to study single-pass absorption values between 2\% and 87\%, respectively.\\

\begin{figure*}[ht]
\centering
\includegraphics[width=0.5\textwidth]{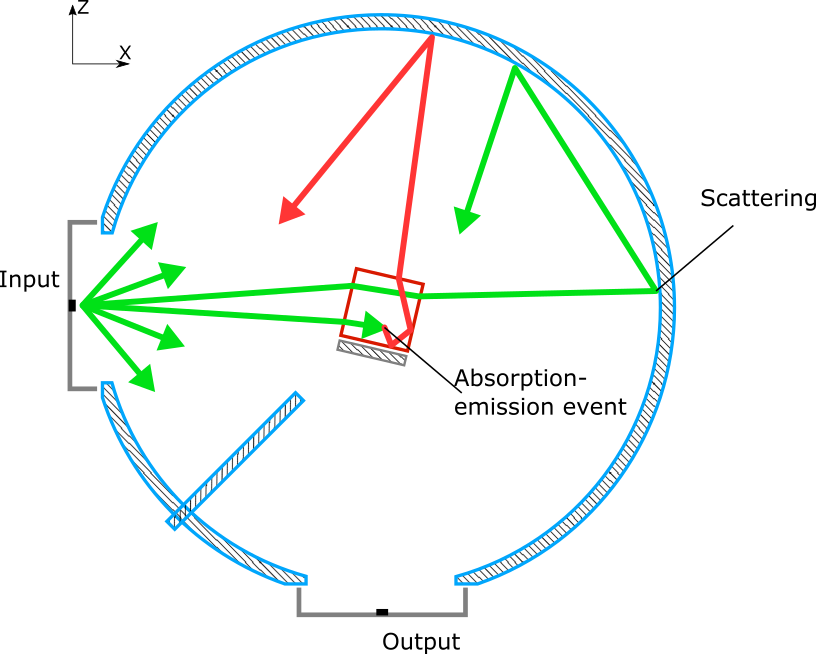}
\caption{Possible paths in the IS for excitation photons (green) which are converted to emission photons (red).}
\label{figureS5}
\end{figure*}

An example of the path of an excitation photon is depicted in Fig.~\ref{figureS5}. A fraction of the initial photon flux that enters the sphere via the input port can, after a series of scattering events on various surfaces, end up in the active volume of the sample. There we sample the possibility of a photon being absorbed along its way across the volume (according to Lambert-Beer law) or leaving the volume unaffected. If absorption occurs, we sample the possibility that the photon is (re-)emitted with a probability given by the PL QY. Emission photons are emitted in a random direction uniformly distributed from the point of absorption. We do not consider re-absorption, which has been discussed in detail elsewhere \cite{Ahn2007}, by setting $A(\lambda_{em})$ = 0. Photons of both kinds (i.e. excitation and emission) scatter around in the sphere, but also can be absorbed by loss channels such as the input port (modelled as Lambertian surfaces with 0 reflectance) or by the IS coating. Paths are terminated when they hit such a loss channel. Moreover to prevent photon paths of infinite lengths, we use the so called Russian roulette technique \cite{Dutre2002,Pharr2010} to stochastically terminate the simulation of photons with some probability after each interaction. In case the photon’s path is not terminated its energy weight is proportionally increased in order to compensate for terminated photons in such a way that the expected result of the simulation stays unaffected.\\

\begin{figure*}[ht]
\centering
\includegraphics[width=0.9\textwidth]{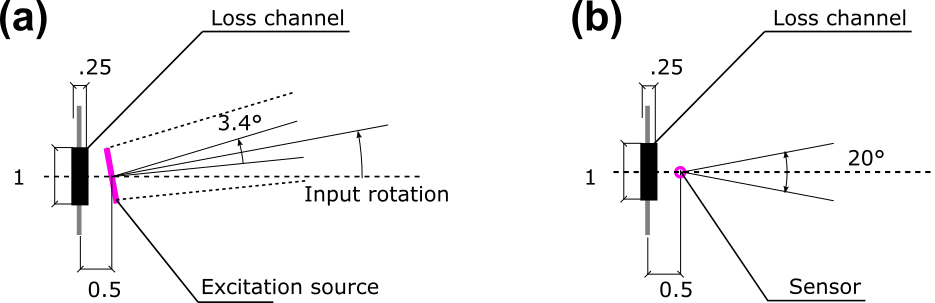}
\caption{Input (a) and output (b) port geometries. The basic unit is derived from the size of an optical fiber with a diameter of 1 mm. The loss channels are modelled as a small rectangular geometry of zero reflectance. The input geometry (a) is capable of rotation around Y-axis while angle of emission is fixed and set to 3.4 degrees. The output geometry (b) cannot be rotated and acceptance angle is set to 20 degrees.}
\label{figureS6}
\end{figure*}

The input and output port geometries containing the excitation source and sensor are depicted in Fig.~\ref{figureS6}. Since in reality, light can exit the sphere via these ports, we insert in our model loss channels behind the excitation source and sensor so that photons hitting these are also removed from the IS. The loss channels are modelled as small rectangular geometries with zero reflectance sitting on aluminum beds. The excitation source in the input port is represented as a small rectangular region in space without actual visible geometry associated. Each point of that region sends out photons into a cone with angle of 3.4 degrees. The sensor in the output port is modelled as a pinhole camera without any associated geometry either, with a field of view equal to 20 degrees, which approximates the optical fiber with NA = 0.22 used in our experiments.\\

Within the cone set by the NA, the sensor measures the flux impinging on it from different directions. This cone is divided into a disjunct set of cells (“pixels”) each of which accumulates weights of excitation and emission photons coming from the associated directions. The resulting accumulated values in all cells are regarded as realizations of the same underlying random variable for which we estimate the mean and variance. They represent the incoming flux on the sensor, which is directly analogous to the measured photons count in the real experiment. For estimating the QY we follow the methodology of the physical experiment described in Eq.~\ref{eq1} (main text) using a test and a reference sample, where the latter does not absorb.\\

The cuvette in the setup is modeled according to a typical commercially available cuvette (Hellma Analytics). There are several material interfaces (Air/Glass, Glass/Air, Glass/Liquid, Air/Liquid) which are accounted for in order to achieve appropriate accuracy of our model (depicted in Fig.~\ref{figureS1}c). The cap is modeled as Lambertian diffuser with the reflectance of 0.9.\\

The cuvette holder is made of aluminum and serves also as a baffle preventing light going from cuvette directly towards output port. Reflectance of aluminum at $\lambda_{exc}$ and $\lambda_{em}$ wavelengths is computed from Fresnel equations.

\subsection*{Analytical model}

\begin{figure*}[ht]
\centering
\includegraphics[width=0.75\textwidth]{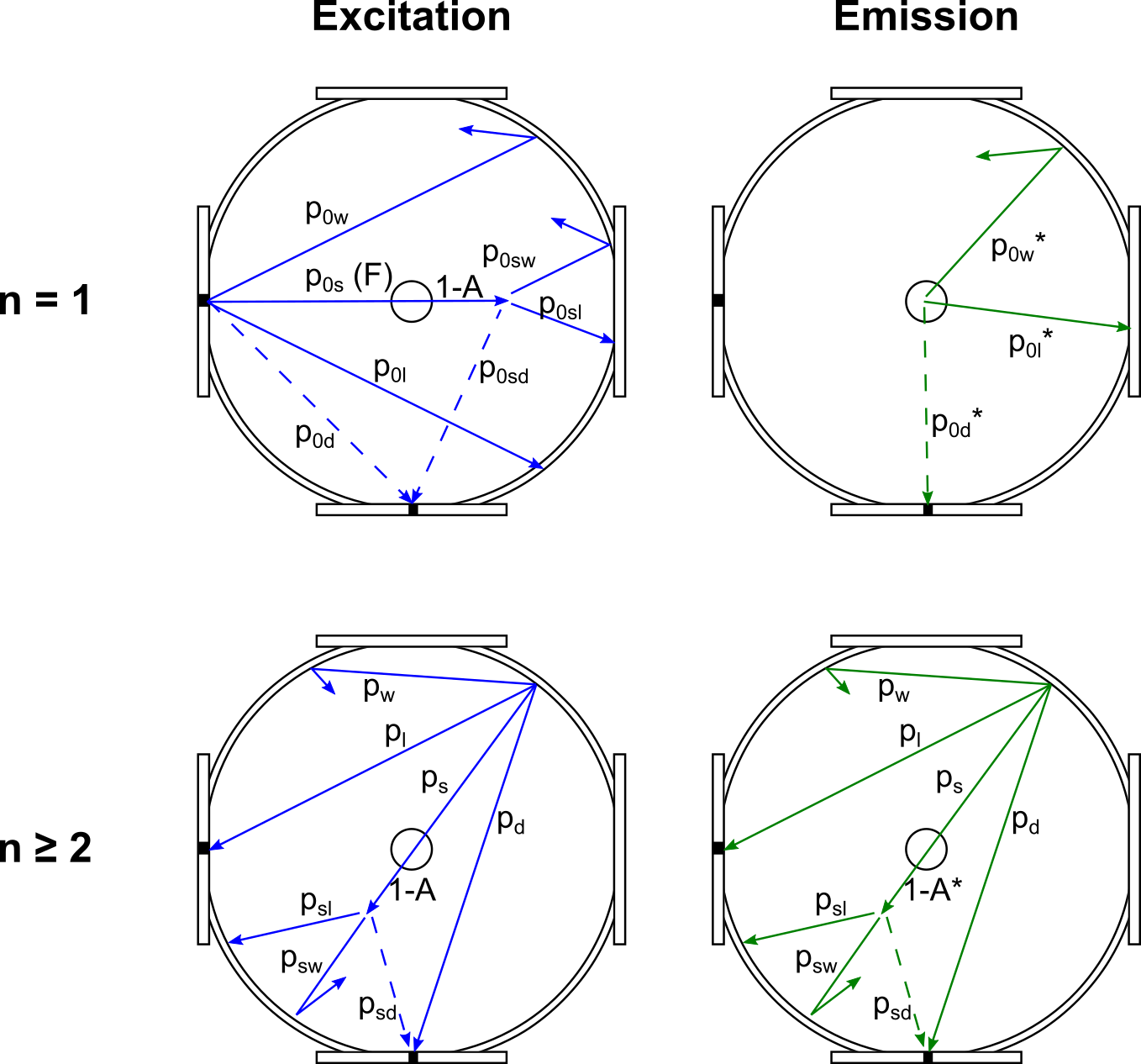}
\caption{Schematic of a generalized IS setup. Lines represent the different pathways for excitation (blue) and emission (green) light between different objects inside the IS: wall (w), loss channel (l), detector (d) and sample (s); $p_{xy}$ represents the probability of going from x to y. Light paths shown by dashed lines are prevented by baffles. }
\label{figureS7}
\end{figure*}

In this approach we simplify the setup illustrated in Figure~\ref{figureS1} by assuming a general IS geometry as shown in Fig.~\ref{figureS7} and described in detail in Ref.~\cite{vanDam2018}. Within this geometry we simulate the detected and absorbed intensities after each consecutive reflection of the excitation and emission light, where we separate the first reflection (n=1) from the consecutive ones (n>1). For n>1, we assume that light reflected from the different areas of the integrating sphere wall has the same probability of hitting an object in the IS, which we set to the relative area of the object to the area of IS interior. Furthermore we assume that the reference sample does not absorb, that the IS is coated with an ideal diffuse light scatterer with a reflectivity of 0.97 and 0.99 at the excitation and emission wavelength respectively and that light can be detected only after the second reflection only due to the presence of baffles. For detailed equations, we refer to Ref.~\cite{vanDam2018}. Simulations were carried out for an IS with a diameter of 10 cm, a spherical sample with a diameter of 1 cm and in- and output ports with a diameter of 4 and 1 mm respectively. 
To account for measurement uncertainty we add noise in to our simulations by describing the measured photon intensities predicted by the AM~\cite{vanDam2018} by a normal distribution. By drawing semi-randomly from this distribution we obtain the probability distribution of the simulated QY. The mode of this distribution is taken to be the most-likely measured QY. For more details we refer to Ref.~\cite{vanDam2018}. \\

\begin{figure*}[ht]
\centering
\includegraphics[width=\textwidth]{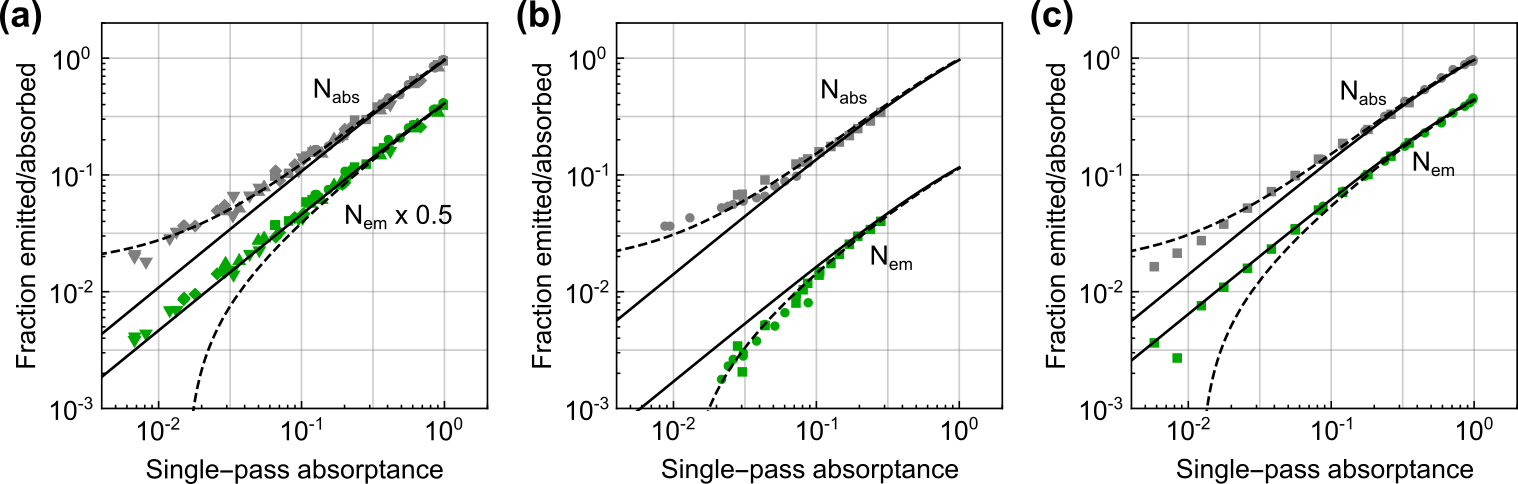}
\caption{Fraction of emitted and absorbed photons plotted as a function of the single-pass absorptance for (a) R6G, (b) CdSe QDs and (c) Si QDs. Solid black lines are fit using analytical model (AM) approach.}
\label{figureS8}
\end{figure*}

\section*{REFERENCES}


\begin{thebibliography}{42}%
\makeatletter
\providecommand \@ifxundefined [1]{%
 \@ifx{#1\undefined}
}%
\providecommand \@ifnum [1]{%
 \ifnum #1\expandafter \@firstoftwo
 \else \expandafter \@secondoftwo
 \fi
}%
\providecommand \@ifx [1]{%
 \ifx #1\expandafter \@firstoftwo
 \else \expandafter \@secondoftwo
 \fi
}%
\providecommand \natexlab [1]{#1}%
\providecommand \enquote  [1]{``#1''}%
\providecommand \bibnamefont  [1]{#1}%
\providecommand \bibfnamefont [1]{#1}%
\providecommand \citenamefont [1]{#1}%
\providecommand \href@noop [0]{\@secondoftwo}%
\providecommand \href [0]{\begingroup \@sanitize@url \@href}%
\providecommand \@href[1]{\@@startlink{#1}\@@href}%
\providecommand \@@href[1]{\endgroup#1\@@endlink}%
\providecommand \@sanitize@url [0]{\catcode `\\12\catcode `\$12\catcode
  `\&12\catcode `\#12\catcode `\^12\catcode `\_12\catcode `\%12\relax}%
\providecommand \@@startlink[1]{}%
\providecommand \@@endlink[0]{}%
\providecommand \url  [0]{\begingroup\@sanitize@url \@url }%
\providecommand \@url [1]{\endgroup\@href {#1}{\urlprefix }}%
\providecommand \urlprefix  [0]{URL }%
\providecommand \Eprint [0]{\href }%
\providecommand \doibase [0]{http://dx.doi.org/}%
\providecommand \selectlanguage [0]{\@gobble}%
\providecommand \bibinfo  [0]{\@secondoftwo}%
\providecommand \bibfield  [0]{\@secondoftwo}%
\providecommand \translation [1]{[#1]}%
\providecommand \BibitemOpen [0]{}%
\providecommand \bibitemStop [0]{}%
\providecommand \bibitemNoStop [0]{.\EOS\space}%
\providecommand \EOS [0]{\spacefactor3000\relax}%
\providecommand \BibitemShut  [1]{\csname bibitem#1\endcsname}%
\let\auto@bib@innerbib\@empty
\bibitem [{\citenamefont {W{\"{u}}rth}\ \emph {et~al.}(2015)\citenamefont
  {W{\"{u}}rth}, \citenamefont {Gei{\ss}ler}, \citenamefont {Behnke},
  \citenamefont {Kaiser},\ and\ \citenamefont {Resch-Genger}}]{Wurth2015}%
  \BibitemOpen
  \bibfield  {author} {\bibinfo {author} {\bibfnamefont {C.}~\bibnamefont
  {W{\"{u}}rth}}, \bibinfo {author} {\bibfnamefont {D.}~\bibnamefont
  {Gei{\ss}ler}}, \bibinfo {author} {\bibfnamefont {T.}~\bibnamefont {Behnke}},
  \bibinfo {author} {\bibfnamefont {M.}~\bibnamefont {Kaiser}}, \ and\ \bibinfo
  {author} {\bibfnamefont {U.}~\bibnamefont {Resch-Genger}},\ }\href {\doibase
  10.1007/s00216-014-8130-z} {\enquote {\bibinfo {title} {{Critical review of
  the determination of photoluminescence quantum yields of luminescent
  reporters}},}\ } (\bibinfo {year} {2015})\BibitemShut {NoStop}%
\bibitem [{\citenamefont {Crosby}\ and\ \citenamefont
  {Demas}(1971)}]{Crosby1971}%
  \BibitemOpen
  \bibfield  {author} {\bibinfo {author} {\bibfnamefont {G.~A.}\ \bibnamefont
  {Crosby}}\ and\ \bibinfo {author} {\bibfnamefont {J.~N.}\ \bibnamefont
  {Demas}},\ }\href {\doibase 10.1021/j100678a001} {\bibfield  {journal}
  {\bibinfo  {journal} {The Journal of Physical Chemistry}\ }\textbf {\bibinfo
  {volume} {75}},\ \bibinfo {pages} {991} (\bibinfo {year} {1971})}\BibitemShut
  {NoStop}%
\bibitem [{\citenamefont {W{\"{u}}rth}\ \emph {et~al.}(2013)\citenamefont
  {W{\"{u}}rth}, \citenamefont {Grabolle}, \citenamefont {Pauli}, \citenamefont
  {Spieles},\ and\ \citenamefont {Resch-Genger}}]{Wurth2013}%
  \BibitemOpen
  \bibfield  {author} {\bibinfo {author} {\bibfnamefont {C.}~\bibnamefont
  {W{\"{u}}rth}}, \bibinfo {author} {\bibfnamefont {M.}~\bibnamefont
  {Grabolle}}, \bibinfo {author} {\bibfnamefont {J.}~\bibnamefont {Pauli}},
  \bibinfo {author} {\bibfnamefont {M.}~\bibnamefont {Spieles}}, \ and\
  \bibinfo {author} {\bibfnamefont {U.}~\bibnamefont {Resch-Genger}},\ }\href
  {\doibase 10.1038/nprot.2013.087} {\bibfield  {journal} {\bibinfo  {journal}
  {Nature Protocols}\ }\textbf {\bibinfo {volume} {8}},\ \bibinfo {pages}
  {1535} (\bibinfo {year} {2013})}\BibitemShut {NoStop}%
\bibitem [{\citenamefont {Valenta}(2014)}]{Valenta2014}%
  \BibitemOpen
  \bibfield  {author} {\bibinfo {author} {\bibfnamefont {J.}~\bibnamefont
  {Valenta}},\ }\href {\doibase 10.1080/21642311.2014.884288} {\bibfield
  {journal} {\bibinfo  {journal} {Nanoscience Methods}\ }\textbf {\bibinfo
  {volume} {3}},\ \bibinfo {pages} {11} (\bibinfo {year} {2014})}\BibitemShut
  {NoStop}%
\bibitem [{\citenamefont {W{\"{u}}rth}\ \emph {et~al.}(2011)\citenamefont
  {W{\"{u}}rth}, \citenamefont {Grabolle}, \citenamefont {Pauli}, \citenamefont
  {Spieles},\ and\ \citenamefont {Resch-Genger}}]{Wurth2011}%
  \BibitemOpen
  \bibfield  {author} {\bibinfo {author} {\bibfnamefont {C.}~\bibnamefont
  {W{\"{u}}rth}}, \bibinfo {author} {\bibfnamefont {M.}~\bibnamefont
  {Grabolle}}, \bibinfo {author} {\bibfnamefont {J.}~\bibnamefont {Pauli}},
  \bibinfo {author} {\bibfnamefont {M.}~\bibnamefont {Spieles}}, \ and\
  \bibinfo {author} {\bibfnamefont {U.}~\bibnamefont {Resch-Genger}},\ }\href
  {\doibase 10.1021/ac2000303} {\bibfield  {journal} {\bibinfo  {journal}
  {Analytical Chemistry}\ }\textbf {\bibinfo {volume} {83}},\ \bibinfo {pages}
  {3431} (\bibinfo {year} {2011})}\BibitemShut {NoStop}%
\bibitem [{\citenamefont {Ahn}\ \emph {et~al.}(2007)\citenamefont {Ahn},
  \citenamefont {Al-Kaysi}, \citenamefont {M{\"{u}}ller}, \citenamefont
  {Wentz},\ and\ \citenamefont {Bardeen}}]{Ahn2007}%
  \BibitemOpen
  \bibfield  {author} {\bibinfo {author} {\bibfnamefont {T.~S.}\ \bibnamefont
  {Ahn}}, \bibinfo {author} {\bibfnamefont {R.~O.}\ \bibnamefont {Al-Kaysi}},
  \bibinfo {author} {\bibfnamefont {A.~M.}\ \bibnamefont {M{\"{u}}ller}},
  \bibinfo {author} {\bibfnamefont {K.~M.}\ \bibnamefont {Wentz}}, \ and\
  \bibinfo {author} {\bibfnamefont {C.~J.}\ \bibnamefont {Bardeen}},\ }\href
  {\doibase 10.1063/1.2768926} {\bibfield  {journal} {\bibinfo  {journal}
  {Review of Scientific Instruments}\ }\textbf {\bibinfo {volume} {78}}
  (\bibinfo {year} {2007}),\ 10.1063/1.2768926}\BibitemShut {NoStop}%
\bibitem [{\citenamefont {Faulkner}\ \emph {et~al.}(2012)\citenamefont
  {Faulkner}, \citenamefont {Mcdowell}, \citenamefont {Price}, \citenamefont
  {Perovic}, \citenamefont {Kherani},\ and\ \citenamefont
  {Ozin}}]{Faulkner2012}%
  \BibitemOpen
  \bibfield  {author} {\bibinfo {author} {\bibfnamefont {D.~O.}\ \bibnamefont
  {Faulkner}}, \bibinfo {author} {\bibfnamefont {J.~J.}\ \bibnamefont
  {Mcdowell}}, \bibinfo {author} {\bibfnamefont {A.~J.}\ \bibnamefont {Price}},
  \bibinfo {author} {\bibfnamefont {D.~D.}\ \bibnamefont {Perovic}}, \bibinfo
  {author} {\bibfnamefont {N.~P.}\ \bibnamefont {Kherani}}, \ and\ \bibinfo
  {author} {\bibfnamefont {G.~A.}\ \bibnamefont {Ozin}},\ }\href {\doibase
  10.1002/lpor.201200077} {\bibfield  {journal} {\bibinfo  {journal} {Laser and
  Photonics Reviews}\ }\textbf {\bibinfo {volume} {6}},\ \bibinfo {pages} {802}
  (\bibinfo {year} {2012})}\BibitemShut {NoStop}%
\bibitem [{\citenamefont {W{\"{u}}rth}\ and\ \citenamefont
  {Resch-Genger}(2015)}]{Wurth2015a}%
  \BibitemOpen
  \bibfield  {author} {\bibinfo {author} {\bibfnamefont {C.}~\bibnamefont
  {W{\"{u}}rth}}\ and\ \bibinfo {author} {\bibfnamefont {U.}~\bibnamefont
  {Resch-Genger}},\ }\href {\doibase 10.1366/14-07679} {\bibfield  {journal}
  {\bibinfo  {journal} {Applied spectroscopy}\ }\textbf {\bibinfo {volume}
  {69}},\ \bibinfo {pages} {749} (\bibinfo {year} {2015})}\BibitemShut
  {NoStop}%
\bibitem [{\citenamefont {Greenham}\ \emph {et~al.}(1995)\citenamefont
  {Greenham}, \citenamefont {Samuel}, \citenamefont {Hayes}, \citenamefont
  {Phillips}, \citenamefont {Kessener}, \citenamefont {Moratti}, \citenamefont
  {Holmes},\ and\ \citenamefont {Friend}}]{Greenham1995}%
  \BibitemOpen
  \bibfield  {author} {\bibinfo {author} {\bibfnamefont {N.~C.}\ \bibnamefont
  {Greenham}}, \bibinfo {author} {\bibfnamefont {I.~D.~W.}\ \bibnamefont
  {Samuel}}, \bibinfo {author} {\bibfnamefont {G.~R.}\ \bibnamefont {Hayes}},
  \bibinfo {author} {\bibfnamefont {R.~T.}\ \bibnamefont {Phillips}}, \bibinfo
  {author} {\bibfnamefont {Y.~A. R.~R.}\ \bibnamefont {Kessener}}, \bibinfo
  {author} {\bibfnamefont {S.~C.}\ \bibnamefont {Moratti}}, \bibinfo {author}
  {\bibfnamefont {A.~B.}\ \bibnamefont {Holmes}}, \ and\ \bibinfo {author}
  {\bibfnamefont {R.~H.}\ \bibnamefont {Friend}},\ }\href {\doibase
  10.1016/0009-2614(95)00584-Q} {\bibfield  {journal} {\bibinfo  {journal}
  {Chemical Physics Letters}\ }\textbf {\bibinfo {volume} {241}},\ \bibinfo
  {pages} {89} (\bibinfo {year} {1995})}\BibitemShut {NoStop}%
\bibitem [{\citenamefont {{De Mello}}, \citenamefont {Wittmann},\ and\
  \citenamefont {Friend}(1997)}]{DeMello1997}%
  \BibitemOpen
  \bibfield  {author} {\bibinfo {author} {\bibfnamefont {J.~C.}\ \bibnamefont
  {{De Mello}}}, \bibinfo {author} {\bibfnamefont {H.~F.}\ \bibnamefont
  {Wittmann}}, \ and\ \bibinfo {author} {\bibfnamefont {R.~H.}\ \bibnamefont
  {Friend}},\ }\href {\doibase 10.1002/adma.19970090308} {\bibfield  {journal}
  {\bibinfo  {journal} {Advanced Materials}\ }\textbf {\bibinfo {volume} {9}},\
  \bibinfo {pages} {230} (\bibinfo {year} {1997})},\ \Eprint
  {http://arxiv.org/abs/97/0302-O23} {arXiv:97/0302-O23 [0935-9648]}
  \BibitemShut {NoStop}%
\bibitem [{\citenamefont {Mangolini}\ \emph {et~al.}(2006)\citenamefont
  {Mangolini}, \citenamefont {Jurbergs}, \citenamefont {Rogojina},\ and\
  \citenamefont {Kortshagen}}]{Mangolini2006}%
  \BibitemOpen
  \bibfield  {author} {\bibinfo {author} {\bibfnamefont {L.}~\bibnamefont
  {Mangolini}}, \bibinfo {author} {\bibfnamefont {D.}~\bibnamefont {Jurbergs}},
  \bibinfo {author} {\bibfnamefont {E.}~\bibnamefont {Rogojina}}, \ and\
  \bibinfo {author} {\bibfnamefont {U.}~\bibnamefont {Kortshagen}},\ }\href
  {\doibase 10.1016/j.jlumin.2006.08.068} {\bibfield  {journal} {\bibinfo
  {journal} {Journal of Luminescence}\ }\textbf {\bibinfo {volume} {121}},\
  \bibinfo {pages} {327} (\bibinfo {year} {2006})}\BibitemShut {NoStop}%
\bibitem [{Hor()}]{Horiba}%
  \BibitemOpen
  \href@noop {} {\enquote {\bibinfo {title}
  {http://www.horiba.com/scientific/marketing-us/quantaphi/},}\ }\BibitemShut
  {NoStop}%
\bibitem [{Ham()}]{Hamamatsu}%
  \BibitemOpen
  \href@noop {} {\enquote {\bibinfo {title}
  {http://www.hamamtsu.com/us/en/product/category/5001/
  5009/5032/c9920-02g/index.html},}\ }\BibitemShut {NoStop}%
\bibitem [{\citenamefont {W{\"{u}}rth}\ \emph {et~al.}(2010)\citenamefont
  {W{\"{u}}rth}, \citenamefont {Lochmann}, \citenamefont {Spieles},
  \citenamefont {Pauli}, \citenamefont {Hoffmann}, \citenamefont
  {Sch{\"{u}}ttrigkeit}, \citenamefont {Franzl},\ and\ \citenamefont
  {Resch-Genger}}]{Wurth2010}%
  \BibitemOpen
  \bibfield  {author} {\bibinfo {author} {\bibfnamefont {C.}~\bibnamefont
  {W{\"{u}}rth}}, \bibinfo {author} {\bibfnamefont {C.}~\bibnamefont
  {Lochmann}}, \bibinfo {author} {\bibfnamefont {M.}~\bibnamefont {Spieles}},
  \bibinfo {author} {\bibfnamefont {J.}~\bibnamefont {Pauli}}, \bibinfo
  {author} {\bibfnamefont {K.}~\bibnamefont {Hoffmann}}, \bibinfo {author}
  {\bibfnamefont {T.}~\bibnamefont {Sch{\"{u}}ttrigkeit}}, \bibinfo {author}
  {\bibfnamefont {T.}~\bibnamefont {Franzl}}, \ and\ \bibinfo {author}
  {\bibfnamefont {U.~T.}\ \bibnamefont {Resch-Genger}},\ }\href {\doibase
  10.1366/000370210791666390} {\bibfield  {journal} {\bibinfo  {journal}
  {Applied Spectroscopy}\ }\textbf {\bibinfo {volume} {64}},\ \bibinfo {pages}
  {733} (\bibinfo {year} {2010})}\BibitemShut {NoStop}%
\bibitem [{\citenamefont {Timmerman}\ \emph {et~al.}(2011)\citenamefont
  {Timmerman}, \citenamefont {Valenta}, \citenamefont {Dohnalov{\'{a}}},
  \citenamefont {{De Boer}},\ and\ \citenamefont
  {Gregorkiewicz}}]{Timmerman2011}%
  \BibitemOpen
  \bibfield  {author} {\bibinfo {author} {\bibfnamefont {D.}~\bibnamefont
  {Timmerman}}, \bibinfo {author} {\bibfnamefont {J.}~\bibnamefont {Valenta}},
  \bibinfo {author} {\bibfnamefont {K.}~\bibnamefont {Dohnalov{\'{a}}}},
  \bibinfo {author} {\bibfnamefont {W.~D.}\ \bibnamefont {{De Boer}}}, \ and\
  \bibinfo {author} {\bibfnamefont {T.}~\bibnamefont {Gregorkiewicz}},\ }\href
  {\doibase 10.1038/nnano.2011.167} {\bibfield  {journal} {\bibinfo  {journal}
  {Nature Nanotechnology}\ }\textbf {\bibinfo {volume} {6}},\ \bibinfo {pages}
  {710} (\bibinfo {year} {2011})}\BibitemShut {NoStop}%
\bibitem [{\citenamefont {Greben}, \citenamefont {Fucikova},\ and\
  \citenamefont {Valenta}(2015)}]{Greben2015}%
  \BibitemOpen
  \bibfield  {author} {\bibinfo {author} {\bibfnamefont {M.}~\bibnamefont
  {Greben}}, \bibinfo {author} {\bibfnamefont {A.}~\bibnamefont {Fucikova}}, \
  and\ \bibinfo {author} {\bibfnamefont {J.}~\bibnamefont {Valenta}},\ }\href
  {\doibase 10.1063/1.4917388} {\bibfield  {journal} {\bibinfo  {journal}
  {Journal of Applied Physics}\ }\textbf {\bibinfo {volume} {117}} (\bibinfo
  {year} {2015}),\ 10.1063/1.4917388}\BibitemShut {NoStop}%
\bibitem [{\citenamefont {Mastronardi}\ \emph {et~al.}(2012)\citenamefont
  {Mastronardi}, \citenamefont {Maier-Flaig}, \citenamefont {Faulkner},
  \citenamefont {Henderson}, \citenamefont {K{\"{u}}bel}, \citenamefont
  {Lemmer},\ and\ \citenamefont {Ozin}}]{Mastronardi2012}%
  \BibitemOpen
  \bibfield  {author} {\bibinfo {author} {\bibfnamefont {M.~L.}\ \bibnamefont
  {Mastronardi}}, \bibinfo {author} {\bibfnamefont {F.}~\bibnamefont
  {Maier-Flaig}}, \bibinfo {author} {\bibfnamefont {D.}~\bibnamefont
  {Faulkner}}, \bibinfo {author} {\bibfnamefont {E.~J.}\ \bibnamefont
  {Henderson}}, \bibinfo {author} {\bibfnamefont {C.}~\bibnamefont
  {K{\"{u}}bel}}, \bibinfo {author} {\bibfnamefont {U.}~\bibnamefont {Lemmer}},
  \ and\ \bibinfo {author} {\bibfnamefont {G.~A.}\ \bibnamefont {Ozin}},\
  }\href {\doibase 10.1021/nl2036194} {\bibfield  {journal} {\bibinfo
  {journal} {Nano Letters}\ }\textbf {\bibinfo {volume} {12}},\ \bibinfo
  {pages} {337} (\bibinfo {year} {2012})}\BibitemShut {NoStop}%
\bibitem [{\citenamefont {Miller}\ \emph {et~al.}(2012)\citenamefont {Miller},
  \citenamefont {{Van Sickle}}, \citenamefont {Anthony}, \citenamefont {Kroll},
  \citenamefont {Kortshagen},\ and\ \citenamefont {Hobbie}}]{Miller2012}%
  \BibitemOpen
  \bibfield  {author} {\bibinfo {author} {\bibfnamefont {J.~B.}\ \bibnamefont
  {Miller}}, \bibinfo {author} {\bibfnamefont {A.~R.}\ \bibnamefont {{Van
  Sickle}}}, \bibinfo {author} {\bibfnamefont {R.~J.}\ \bibnamefont {Anthony}},
  \bibinfo {author} {\bibfnamefont {D.~M.}\ \bibnamefont {Kroll}}, \bibinfo
  {author} {\bibfnamefont {U.~R.}\ \bibnamefont {Kortshagen}}, \ and\ \bibinfo
  {author} {\bibfnamefont {E.~K.}\ \bibnamefont {Hobbie}},\ }\href {\doibase
  10.1021/nn302524k} {\bibfield  {journal} {\bibinfo  {journal} {ACS Nano}\
  }\textbf {\bibinfo {volume} {6}},\ \bibinfo {pages} {7389} (\bibinfo {year}
  {2012})}\BibitemShut {NoStop}%
\bibitem [{\citenamefont {Sun}\ \emph {et~al.}(2015)\citenamefont {Sun},
  \citenamefont {Qian}, \citenamefont {Wang}, \citenamefont {Wei},
  \citenamefont {Mastronardi}, \citenamefont {Casillas}, \citenamefont {Breu},\
  and\ \citenamefont {Ozin}}]{Sun2015}%
  \BibitemOpen
  \bibfield  {author} {\bibinfo {author} {\bibfnamefont {W.}~\bibnamefont
  {Sun}}, \bibinfo {author} {\bibfnamefont {C.}~\bibnamefont {Qian}}, \bibinfo
  {author} {\bibfnamefont {L.}~\bibnamefont {Wang}}, \bibinfo {author}
  {\bibfnamefont {M.}~\bibnamefont {Wei}}, \bibinfo {author} {\bibfnamefont
  {M.~L.}\ \bibnamefont {Mastronardi}}, \bibinfo {author} {\bibfnamefont
  {G.}~\bibnamefont {Casillas}}, \bibinfo {author} {\bibfnamefont
  {J.}~\bibnamefont {Breu}}, \ and\ \bibinfo {author} {\bibfnamefont {G.~A.}\
  \bibnamefont {Ozin}},\ }\href {\doibase 10.1002/adma.201403552} {\bibfield
  {journal} {\bibinfo  {journal} {Advanced Materials}\ }\textbf {\bibinfo
  {volume} {27}},\ \bibinfo {pages} {746} (\bibinfo {year} {2015})}\BibitemShut
  {NoStop}%
\bibitem [{\citenamefont {Kubin}\ and\ \citenamefont
  {Fletcher}(1982)}]{Kubin1982}%
  \BibitemOpen
  \bibfield  {author} {\bibinfo {author} {\bibfnamefont {R.~F.}\ \bibnamefont
  {Kubin}}\ and\ \bibinfo {author} {\bibfnamefont {A.~N.}\ \bibnamefont
  {Fletcher}},\ }\href {\doibase 10.1016/0022-2313(82)90045-X} {\bibfield
  {journal} {\bibinfo  {journal} {Journal of Luminescence}\ }\textbf {\bibinfo
  {volume} {27}},\ \bibinfo {pages} {455} (\bibinfo {year} {1982})},\ \Eprint
  {http://arxiv.org/abs/arXiv:1011.1669v3} {arXiv:arXiv:1011.1669v3}
  \BibitemShut {NoStop}%
\bibitem [{\citenamefont {W{\"{u}}rth}\ \emph {et~al.}(2012)\citenamefont
  {W{\"{u}}rth}, \citenamefont {Gonz{\'{a}}lez}, \citenamefont {Niessner},
  \citenamefont {Panne}, \citenamefont {Haisch},\ and\ \citenamefont
  {Genger}}]{Wurth2012}%
  \BibitemOpen
  \bibfield  {author} {\bibinfo {author} {\bibfnamefont {C.}~\bibnamefont
  {W{\"{u}}rth}}, \bibinfo {author} {\bibfnamefont {M.~G.}\ \bibnamefont
  {Gonz{\'{a}}lez}}, \bibinfo {author} {\bibfnamefont {R.}~\bibnamefont
  {Niessner}}, \bibinfo {author} {\bibfnamefont {U.}~\bibnamefont {Panne}},
  \bibinfo {author} {\bibfnamefont {C.}~\bibnamefont {Haisch}}, \ and\ \bibinfo
  {author} {\bibfnamefont {U.~R.}\ \bibnamefont {Genger}},\ }\href {\doibase
  10.1016/j.talanta.2011.12.051} {\bibfield  {journal} {\bibinfo  {journal}
  {Talanta}\ }\textbf {\bibinfo {volume} {90}},\ \bibinfo {pages} {30}
  (\bibinfo {year} {2012})}\BibitemShut {NoStop}%
\bibitem [{\citenamefont {Wawilow}(1927)}]{Wawilow1927}%
  \BibitemOpen
  \bibfield  {author} {\bibinfo {author} {\bibfnamefont {S.~J.}\ \bibnamefont
  {Wawilow}},\ }\href {\doibase 10.1007/BF01397622} {\bibfield  {journal}
  {\bibinfo  {journal} {Zeitschrift f??r Physik}\ }\textbf {\bibinfo {volume}
  {42}},\ \bibinfo {pages} {311} (\bibinfo {year} {1927})}\BibitemShut
  {NoStop}%
\bibitem [{\citenamefont {Valenta}\ \emph {et~al.}(2014)\citenamefont
  {Valenta}, \citenamefont {Greben}, \citenamefont {Gutsch}, \citenamefont
  {Hiller},\ and\ \citenamefont {Zacharias}}]{Valenta2014a}%
  \BibitemOpen
  \bibfield  {author} {\bibinfo {author} {\bibfnamefont {J.}~\bibnamefont
  {Valenta}}, \bibinfo {author} {\bibfnamefont {M.}~\bibnamefont {Greben}},
  \bibinfo {author} {\bibfnamefont {S.}~\bibnamefont {Gutsch}}, \bibinfo
  {author} {\bibfnamefont {D.}~\bibnamefont {Hiller}}, \ and\ \bibinfo {author}
  {\bibfnamefont {M.}~\bibnamefont {Zacharias}},\ }\href {\doibase
  10.1063/1.4904472} {\bibfield  {journal} {\bibinfo  {journal} {Applied
  Physics Letters}\ }\textbf {\bibinfo {volume} {105}} (\bibinfo {year}
  {2014}),\ 10.1063/1.4904472}\BibitemShut {NoStop}%
\bibitem [{\citenamefont {Saeed}\ \emph {et~al.}(2014)\citenamefont {Saeed},
  \citenamefont {{De Jong}}, \citenamefont {Dohnalova},\ and\ \citenamefont
  {Gregorkiewicz}}]{Saeed2014}%
  \BibitemOpen
  \bibfield  {author} {\bibinfo {author} {\bibfnamefont {S.}~\bibnamefont
  {Saeed}}, \bibinfo {author} {\bibfnamefont {E.~M.}\ \bibnamefont {{De
  Jong}}}, \bibinfo {author} {\bibfnamefont {K.}~\bibnamefont {Dohnalova}}, \
  and\ \bibinfo {author} {\bibfnamefont {T.}~\bibnamefont {Gregorkiewicz}},\
  }\href {\doibase 10.1038/ncomms5665} {\bibfield  {journal} {\bibinfo
  {journal} {Nature Communications}\ }\textbf {\bibinfo {volume} {5}} (\bibinfo
  {year} {2014}),\ 10.1038/ncomms5665}\BibitemShut {NoStop}%
\bibitem [{\citenamefont {Chung}\ \emph {et~al.}(2017)\citenamefont {Chung},
  \citenamefont {Limpens}, \citenamefont {Gregorkiewicz}, \citenamefont {{Xuan
  Chung}}, \citenamefont {Limpens}, \citenamefont {Gregorkiewicz},
  \citenamefont {Chung}, \citenamefont {Limpens},\ and\ \citenamefont
  {Gregorkiewicz}}]{Chung2017}%
  \BibitemOpen
  \bibfield  {author} {\bibinfo {author} {\bibfnamefont {N.~X.}\ \bibnamefont
  {Chung}}, \bibinfo {author} {\bibfnamefont {R.}~\bibnamefont {Limpens}},
  \bibinfo {author} {\bibfnamefont {T.}~\bibnamefont {Gregorkiewicz}}, \bibinfo
  {author} {\bibfnamefont {N.}~\bibnamefont {{Xuan Chung}}}, \bibinfo {author}
  {\bibfnamefont {R.}~\bibnamefont {Limpens}}, \bibinfo {author} {\bibfnamefont
  {T.}~\bibnamefont {Gregorkiewicz}}, \bibinfo {author} {\bibfnamefont {N.~X.}\
  \bibnamefont {Chung}}, \bibinfo {author} {\bibfnamefont {R.}~\bibnamefont
  {Limpens}}, \ and\ \bibinfo {author} {\bibfnamefont {T.}~\bibnamefont
  {Gregorkiewicz}},\ }\href {\doibase 10.1002/adom.201600709} {\bibfield
  {journal} {\bibinfo  {journal} {Advanced Optical Materials}\ }\textbf
  {\bibinfo {volume} {5}} (\bibinfo {year} {2017}),\
  10.1002/adom.201600709}\BibitemShut {NoStop}%
\bibitem [{\citenamefont {Miura}\ \emph {et~al.}(2006)\citenamefont {Miura},
  \citenamefont {Nakamura}, \citenamefont {Fujii}, \citenamefont {Inui},\ and\
  \citenamefont {Hayashi}}]{Miura2006}%
  \BibitemOpen
  \bibfield  {author} {\bibinfo {author} {\bibfnamefont {S.}~\bibnamefont
  {Miura}}, \bibinfo {author} {\bibfnamefont {T.}~\bibnamefont {Nakamura}},
  \bibinfo {author} {\bibfnamefont {M.}~\bibnamefont {Fujii}}, \bibinfo
  {author} {\bibfnamefont {M.}~\bibnamefont {Inui}}, \ and\ \bibinfo {author}
  {\bibfnamefont {S.}~\bibnamefont {Hayashi}},\ }\href {\doibase
  10.1103/PhysRevB.73.245333} {\bibfield  {journal} {\bibinfo  {journal}
  {Physical Review B}\ }\textbf {\bibinfo {volume} {73}},\ \bibinfo {pages}
  {245333} (\bibinfo {year} {2006})}\BibitemShut {NoStop}%
\bibitem [{\citenamefont {Limpens}\ \emph {et~al.}(2015)\citenamefont
  {Limpens}, \citenamefont {Lesage}, \citenamefont {Stallinga}, \citenamefont
  {Poddubny}, \citenamefont {Fujii},\ and\ \citenamefont
  {Gregorkiewicz}}]{Limpens2015}%
  \BibitemOpen
  \bibfield  {author} {\bibinfo {author} {\bibfnamefont {R.}~\bibnamefont
  {Limpens}}, \bibinfo {author} {\bibfnamefont {A.}~\bibnamefont {Lesage}},
  \bibinfo {author} {\bibfnamefont {P.}~\bibnamefont {Stallinga}}, \bibinfo
  {author} {\bibfnamefont {A.~N.}\ \bibnamefont {Poddubny}}, \bibinfo {author}
  {\bibfnamefont {M.}~\bibnamefont {Fujii}}, \ and\ \bibinfo {author}
  {\bibfnamefont {T.}~\bibnamefont {Gregorkiewicz}},\ }\href {\doibase
  10.1021/acs.jpcc.5b06339} {\bibfield  {journal} {\bibinfo  {journal} {Journal
  of Physical Chemistry C}\ }\textbf {\bibinfo {volume} {119}},\ \bibinfo
  {pages} {19565} (\bibinfo {year} {2015})}\BibitemShut {NoStop}%
\bibitem [{\citenamefont {van Dam}\ \emph {et~al.}(2018)\citenamefont {van
  Dam}, \citenamefont {Dohnal}, \citenamefont {Bruhn},\ and\ \citenamefont
  {Dohnalov\'{a}}}]{vanDam2018}%
  \BibitemOpen
  \bibfield  {author} {\bibinfo {author} {\bibfnamefont {B.}~\bibnamefont {van
  Dam}}, \bibinfo {author} {\bibfnamefont {G.}~\bibnamefont {Dohnal}}, \bibinfo
  {author} {\bibfnamefont {B.}~\bibnamefont {Bruhn}}, \ and\ \bibinfo {author}
  {\bibfnamefont {K.}~\bibnamefont {Dohnalov\'{a}}},\ }\href@noop {} {\bibfield
   {journal} {\bibinfo  {journal} {arXiv:1711.06200 [physics.ins-det], to
  appear in AIP Advances 2018}\ } (\bibinfo {year} {2018})}\BibitemShut
  {NoStop}%
\bibitem [{\citenamefont {Cedilnik}, \citenamefont {Kosmelj},\ and\
  \citenamefont {Blejec}(2004)}]{blejec}%
  \BibitemOpen
  \bibfield  {author} {\bibinfo {author} {\bibfnamefont {A.}~\bibnamefont
  {Cedilnik}}, \bibinfo {author} {\bibfnamefont {K.}~\bibnamefont {Kosmelj}}, \
  and\ \bibinfo {author} {\bibfnamefont {A.}~\bibnamefont {Blejec}},\
  }\href@noop {} {\bibfield  {journal} {\bibinfo  {journal} {Metodoloski
  zvezki}\ }\textbf {\bibinfo {volume} {1}},\ \bibinfo {pages} {99} (\bibinfo
  {year} {2004})}\BibitemShut {NoStop}%
\bibitem [{\citenamefont {Grabolle}\ \emph {et~al.}(2009)\citenamefont
  {Grabolle}, \citenamefont {Spieles}, \citenamefont {Lesnyak}, \citenamefont
  {Gaponik}, \citenamefont {Eychm{\"{u}}ller},\ and\ \citenamefont
  {Resch-Genger}}]{Grabolle2009}%
  \BibitemOpen
  \bibfield  {author} {\bibinfo {author} {\bibfnamefont {M.}~\bibnamefont
  {Grabolle}}, \bibinfo {author} {\bibfnamefont {M.}~\bibnamefont {Spieles}},
  \bibinfo {author} {\bibfnamefont {V.}~\bibnamefont {Lesnyak}}, \bibinfo
  {author} {\bibfnamefont {N.}~\bibnamefont {Gaponik}}, \bibinfo {author}
  {\bibfnamefont {A.}~\bibnamefont {Eychm{\"{u}}ller}}, \ and\ \bibinfo
  {author} {\bibfnamefont {U.}~\bibnamefont {Resch-Genger}},\ }\href {\doibase
  10.1021/ac900308v} {\bibfield  {journal} {\bibinfo  {journal} {Analytical
  Chemistry}\ }\textbf {\bibinfo {volume} {81}},\ \bibinfo {pages} {6285}
  (\bibinfo {year} {2009})}\BibitemShut {NoStop}%
\bibitem [{\citenamefont {{De Roo}}\ \emph {et~al.}(2016)\citenamefont {{De
  Roo}}, \citenamefont {Ib{\'{a}}{\~{n}}ez}, \citenamefont {Geiregat},
  \citenamefont {Nedelcu}, \citenamefont {Walravens}, \citenamefont {Maes},
  \citenamefont {Martins}, \citenamefont {{Van Driessche}}, \citenamefont
  {Kovalenko},\ and\ \citenamefont {Hens}}]{DeRoo2016}%
  \BibitemOpen
  \bibfield  {author} {\bibinfo {author} {\bibfnamefont {J.}~\bibnamefont {{De
  Roo}}}, \bibinfo {author} {\bibfnamefont {M.}~\bibnamefont
  {Ib{\'{a}}{\~{n}}ez}}, \bibinfo {author} {\bibfnamefont {P.}~\bibnamefont
  {Geiregat}}, \bibinfo {author} {\bibfnamefont {G.}~\bibnamefont {Nedelcu}},
  \bibinfo {author} {\bibfnamefont {W.}~\bibnamefont {Walravens}}, \bibinfo
  {author} {\bibfnamefont {J.}~\bibnamefont {Maes}}, \bibinfo {author}
  {\bibfnamefont {J.~C.}\ \bibnamefont {Martins}}, \bibinfo {author}
  {\bibfnamefont {I.}~\bibnamefont {{Van Driessche}}}, \bibinfo {author}
  {\bibfnamefont {M.~V.}\ \bibnamefont {Kovalenko}}, \ and\ \bibinfo {author}
  {\bibfnamefont {Z.}~\bibnamefont {Hens}},\ }\href {\doibase
  10.1021/acsnano.5b06295} {\bibfield  {journal} {\bibinfo  {journal} {ACS
  Nano}\ }\textbf {\bibinfo {volume} {10}},\ \bibinfo {pages} {2071} (\bibinfo
  {year} {2016})}\BibitemShut {NoStop}%
\bibitem [{\citenamefont {Semonin}\ \emph {et~al.}(2010)\citenamefont
  {Semonin}, \citenamefont {Johnson}, \citenamefont {Luther}, \citenamefont
  {Midgett}, \citenamefont {Nozik},\ and\ \citenamefont {Beard}}]{Semonin2010}%
  \BibitemOpen
  \bibfield  {author} {\bibinfo {author} {\bibfnamefont {O.~E.}\ \bibnamefont
  {Semonin}}, \bibinfo {author} {\bibfnamefont {J.~C.}\ \bibnamefont
  {Johnson}}, \bibinfo {author} {\bibfnamefont {J.~M.}\ \bibnamefont {Luther}},
  \bibinfo {author} {\bibfnamefont {A.~G.}\ \bibnamefont {Midgett}}, \bibinfo
  {author} {\bibfnamefont {A.~J.}\ \bibnamefont {Nozik}}, \ and\ \bibinfo
  {author} {\bibfnamefont {M.~C.}\ \bibnamefont {Beard}},\ }\href {\doibase
  10.1021/jz100830r} {\bibfield  {journal} {\bibinfo  {journal} {Journal of
  Physical Chemistry Letters}\ }\textbf {\bibinfo {volume} {1}},\ \bibinfo
  {pages} {2445} (\bibinfo {year} {2010})}\BibitemShut {NoStop}%
\bibitem [{\citenamefont {Birge}(1987)}]{Birge1987}%
  \BibitemOpen
  \bibfield  {author} {\bibinfo {author} {\bibfnamefont {R.}~\bibnamefont
  {Birge}},\ }\href@noop {} {\bibfield  {journal} {\bibinfo  {journal} {Kodak
  Publ.}\ ,\ \bibinfo {pages} {JJ}} (\bibinfo {year} {1987})}\BibitemShut
  {NoStop}%
\bibitem [{\citenamefont {Kelly}\ and\ \citenamefont
  {Veinot}(2010)}]{Kelly2010}%
  \BibitemOpen
  \bibfield  {author} {\bibinfo {author} {\bibfnamefont {J.~A.}\ \bibnamefont
  {Kelly}}\ and\ \bibinfo {author} {\bibfnamefont {J.~G.~C.}\ \bibnamefont
  {Veinot}},\ }\href {\doibase 10.1021/nn101022b} {\bibfield  {journal}
  {\bibinfo  {journal} {ACS Nano}\ }\textbf {\bibinfo {volume} {4}},\ \bibinfo
  {pages} {4645} (\bibinfo {year} {2010})}\BibitemShut {NoStop}%
\bibitem [{\citenamefont {Hessel}, \citenamefont {Henderson},\ and\
  \citenamefont {Veinot}(2006)}]{Hessel2006}%
  \BibitemOpen
  \bibfield  {author} {\bibinfo {author} {\bibfnamefont {C.~M.}\ \bibnamefont
  {Hessel}}, \bibinfo {author} {\bibfnamefont {E.~J.}\ \bibnamefont
  {Henderson}}, \ and\ \bibinfo {author} {\bibfnamefont {J.~G.}\ \bibnamefont
  {Veinot}},\ }\href {\doibase 10.1021/cm0602803} {\bibfield  {journal}
  {\bibinfo  {journal} {Chemistry of Materials}\ }\textbf {\bibinfo {volume}
  {18}},\ \bibinfo {pages} {6139} (\bibinfo {year} {2006})}\BibitemShut
  {NoStop}%
\bibitem [{\citenamefont {Valenta}\ \emph {et~al.}(2016)\citenamefont
  {Valenta}, \citenamefont {Greben}, \citenamefont {Reme{\v{s}}}, \citenamefont
  {Gutsch}, \citenamefont {Hiller},\ and\ \citenamefont
  {Zacharias}}]{Valenta2016}%
  \BibitemOpen
  \bibfield  {author} {\bibinfo {author} {\bibfnamefont {J.}~\bibnamefont
  {Valenta}}, \bibinfo {author} {\bibfnamefont {M.}~\bibnamefont {Greben}},
  \bibinfo {author} {\bibfnamefont {Z.}~\bibnamefont {Reme{\v{s}}}}, \bibinfo
  {author} {\bibfnamefont {S.}~\bibnamefont {Gutsch}}, \bibinfo {author}
  {\bibfnamefont {D.}~\bibnamefont {Hiller}}, \ and\ \bibinfo {author}
  {\bibfnamefont {M.}~\bibnamefont {Zacharias}},\ }\href {\doibase
  10.1063/1.4939699} {\bibfield  {journal} {\bibinfo  {journal} {Applied
  Physics Letters}\ }\textbf {\bibinfo {volume} {108}} (\bibinfo {year}
  {2016}),\ 10.1063/1.4939699}\BibitemShut {NoStop}%
\bibitem [{\citenamefont {Chung}, \citenamefont {Limpens},\ and\ \citenamefont
  {Gregorkiewicz}(2015)}]{Chung2015}%
  \BibitemOpen
  \bibfield  {author} {\bibinfo {author} {\bibfnamefont {N.}~\bibnamefont
  {Chung}}, \bibinfo {author} {\bibfnamefont {R.}~\bibnamefont {Limpens}}, \
  and\ \bibinfo {author} {\bibfnamefont {T.}~\bibnamefont {Gregorkiewicz}},\
  }in\ \href {\doibase 10.1117/12.2191105} {\emph {\bibinfo {booktitle}
  {Proceedings of SPIE - The International Society for Optical Engineering}}},\
  Vol.\ \bibinfo {volume} {9562}\ (\bibinfo {year} {2015})\BibitemShut
  {NoStop}%
\bibitem [{\citenamefont {Duyens}(1956)}]{Duyens1956}%
  \BibitemOpen
  \bibfield  {author} {\bibinfo {author} {\bibfnamefont {L.}~\bibnamefont
  {Duyens}},\ }\href {\doibase 10.1016/0006-3002(56)90380-8} {\bibfield
  {journal} {\bibinfo  {journal} {Biochimica et Biophysica Acta}\ }\textbf
  {\bibinfo {volume} {19}},\ \bibinfo {pages} {1} (\bibinfo {year}
  {1956})}\BibitemShut {NoStop}%
\bibitem [{\citenamefont {Chandrasekhar}(1960)}]{Chandrasekhar1960}%
  \BibitemOpen
  \bibfield  {author} {\bibinfo {author} {\bibfnamefont {S.}~\bibnamefont
  {Chandrasekhar}},\ }\href {\doibase 10.1007/SpringerReference_221860}
  {\bibfield  {journal} {\bibinfo  {journal} {New York Dover}\ ,\ \bibinfo
  {pages} {393}} (\bibinfo {year} {1960})}\BibitemShut {NoStop}%
\bibitem [{\citenamefont {Dutr{\'{e}}}, \citenamefont {Bala},\ and\
  \citenamefont {Bekaert}(2002)}]{Dutre2002}%
  \BibitemOpen
  \bibfield  {author} {\bibinfo {author} {\bibfnamefont {P.}~\bibnamefont
  {Dutr{\'{e}}}}, \bibinfo {author} {\bibfnamefont {K.}~\bibnamefont {Bala}}, \
  and\ \bibinfo {author} {\bibfnamefont {P.}~\bibnamefont {Bekaert}},\ }\href
  {\doibase 10.1017/CBO9781107415324.004} {\bibfield  {journal} {\bibinfo
  {journal} {Odonatologica}\ }\textbf {\bibinfo {volume} {44}},\ \bibinfo
  {pages} {447} (\bibinfo {year} {2002})},\ \Eprint
  {http://arxiv.org/abs/arXiv:1011.1669v3} {arXiv:arXiv:1011.1669v3}
  \BibitemShut {NoStop}%
\bibitem [{\citenamefont {Veach}(1997)}]{Veach1997}%
  \BibitemOpen
  \bibfield  {author} {\bibinfo {author} {\bibfnamefont {E.}~\bibnamefont
  {Veach}},\ }\href {\doibase 10.1016/S1350-9462(98)00033-0} {\bibfield
  {journal} {\bibinfo  {journal} {Dissertation at the Department of Computer
  Science of Stanford University}\ }\textbf {\bibinfo {volume} {134}},\
  \bibinfo {pages} {759} (\bibinfo {year} {1997})}\BibitemShut {NoStop}%
\bibitem [{\citenamefont {Pharr}\ and\ \citenamefont
  {Humphreys}(2010)}]{Pharr2010}%
  \BibitemOpen
  \bibfield  {author} {\bibinfo {author} {\bibfnamefont {M.}~\bibnamefont
  {Pharr}}\ and\ \bibinfo {author} {\bibfnamefont {G.}~\bibnamefont
  {Humphreys}},\ }\href {\doibase 10.1016/B978-0-12-375079-2.50001-9} {\emph
  {\bibinfo {title} {Physically Based Rendering}}}\ (\bibinfo {year} {2010})\
  pp.\ \bibinfo {pages} {1--52}\BibitemShut {NoStop}%
\end{thebibliography}

%

\end{document}